\documentclass[twocolumn,runningheads]{svjour2}

\smartqed  

\usepackage{graphicx}
\usepackage[T1]{fontenc}
\usepackage[latin1]{inputenc}
\usepackage{multirow}
\usepackage{supertabular}
\usepackage{color}

\begin{document}

\journalname{International Journal of Advanced Manufacturing Technology} 

\title{Self-excited vibrations in turning : Forces torsor analysis}


\author{Claudiu F. Bisu \and Alain~G\'erard \and Jean-Yves~K'nevez \and Raynald~Laheurte \and Olivier~Cahuc}

\institute{C.F. Bisu \at University Politehnica from Bucharest,\\313 Splaiul Independentei,  060042 Bucharest Roumanie (UE)
              \\\email{cfbisu@gmail.com}           
           \and A. G\'erard (corresponding author) \and C. F. Bisu \and O. Cahuc \and J-Y. K'nevez \and R. Laheurte \at
              Universit\'e de Bordeaux,\\351 cours de la Lib\'eration, 33405 Talence-cedex France (UE)\\
              Tel.: +33 (0)5 40 00 62 23\\
              Fax: +33 (0)5 40 00 69 64
              \\\email{alain.gerard@u-bordeaux1.fr}   
\and R. Laheurte \at Universit\'e de Bordeaux - IUT EA 496,\\15 rue Naudet, 33175 Gradignan Cedex France (UE) \\\email{raynald.laheurte@u-bordeaux1.fr} }

\date{Received: date / Accepted: date}

\maketitle

\begin{abstract}
The present work deals with determining the necessary parameters considering a three dimensional model to simulate in a realistic way the turning process on machine tool. 

This paper is dedicated to the study of the self-excited vibrations incidence on various major mechanics characteristics of the system workpiece / tool / material. The efforts (forces and moments) measurement using a six components dynamometer confirms the tool tip moments existence. 

The fundamental frequency of 190~Hz proves to be common to the tool tip point displacements, the action application point or at the torque exerted to the tool tip point. The confrontation of the results concerning displacements and efforts shows that the applications points of these elements evolve according to similar ellipses located in quasi identical planes. The large and the small axes of these ellipses are increasing with the feed rate motion values accordingly to the mechanical power injected into the system. Conversely, the respective axes ratios of these ellipses are decreasing functions of the feed rate while the ratio of these ratios remains constant when the feed rate value is increasing.

In addition, some chip characteristics are given, like the thickness variations, the width or the hardening phenomenon.
	

\end{abstract}

\section*{Nomenclature}
\label{sec:Nomenclature}

	\begin{tabular}{lp{6cm}}
\raggedright 

A & Central axis point;\\
$[A]_{o}$ & Actions torsor exerted on the tool tip in O point;  \\
$ap$ & Cutting depth (mm); \\
$a_{f}$ ($b_{f}$) & Large (small) ellipse axis attached to the geometrical place of forces application points;\\
\textbf{BT} & Block Tool; \\
\textbf{BW} & Block Workpiece; \\
$d_{i}$ & Directory line of moments projection to the central axis (i = 1, 2); \\
$e_{i}$ & Ellipse point belong attached of forces application points (i=1-5); \\
$f$  & Feed rate (mm/rev); \\
$f_{cop}$ & Chip segmentation frequency (Hz);\\
$F_{v}$ ($F_{n}$) & Variable (nominal) cutting force (N);\\
$F_{x}$ & Effort on cross direction (N); \\
$F_{y}$ & Effort on cutting axis (N); \\
$F_{z}$ & Effort on feed rate axis (N); \\
$h_{max}$ & Maximum chip thickness (mm); \\
$h_{min}$ & Minimum chip thickness (mm); \\
$l_{0}$ & Chip undulation length (mm); \\
$M_{_{A}}$ & Cutting forces minimum moment in A who actuate on the tool (N.m);\\
$M_{_{O}}$ & Cutting forces moment in the O point who actuate on the tool (N.m); \\
$N$ & Spindle speed (rpm); \\
$n_{f}$ & $P_{f}$ normal direction; \\
$n_{fa}$ ($n_{fb}$) & Normal direction projection to the $P_{f}$ on the $a_{f}$ ($b_{f}$); \\
O & Tool point reference; \\
O' & Dynamometer center transducer; \\
$P_{f}$ & Plane attached to the forces application points; \\

\end{tabular}

\section*{Nomenclature continuation}
\label{sec:Nomenclature2}

	\begin{tabular}{lp{6.8cm}}
\raggedright

$P_{u}$ & Plane attached to the tool point displacements; \\
$R$ & Cutting forces vector sum who actuate on the tool (N); \\
$r_{\epsilon}$ & Cutting edge radius (mm); \\
R & Sharpness radius (mm); \\
$R_{t}$ & Roughness $(\mu m)$; \\
$ T $ & Time related to one revolution of workpiece (s); \\
u & Tool tip point displacement (m); \\
$V$ & Cutting speed (m/min); \\
$w_{max}$ & Maximum chip width (mm); \\
$w_{min}$ & Minimum chip width (mm); \\
\textbf{WTM} & Workpiece-Tool-Machine tool; \\
x & Radial direction; \\
y	&	Cutting axis; \\
z	&	Feed rate direction; \\
$\alpha$	&	Clearance angle (degree); \\
$\alpha_{\kappa(xy)}$	&	Angle of main tool displacements direction included in the plane (x,y) (degree); \\
$\alpha_{\kappa(yz)}$	&	Angle of main tool displacements direction included in the plane (y,z) (degree); \\
$\Delta t$	&	Time corresponding of phase difference between two signals (s);\\
$\Phi$	&	Primary share angle (degree); \\
$\varphi _{c}$	&	Chip width slopes angle between each undulation (degree); \\
$\varphi _{fu_{i}}$	&	Phase difference between the tool tip displacements components i=x, y, z (degree); \\
$\gamma$	&	Cutting angle (degree); \\
$\lambda_{s}$	&	Inclination angle of tool edge (degree); \\
$\kappa_{r}$	&	Direct angle (degree); \\
$\theta_{e(xy)}$	&	Stiffness principal direction angle related to the plane (x,y) (degree); \\
$\theta_{e(yz)}$	&	Stiffness principal direction angle related to the plane (y,z) (degree); \\
$\xi_{c}$	&	Chip hardening coefficient; \\
\end{tabular}

\section{Introduction}
\label{sec:1}

In the three-dimensional cutting case, the mechanical actions torsor (forces and moments), is often truncated: the moments part of this torsor is neglected fault of adapted metrology  \cite{mehdi-A-play-02b}. However, efforts and  pure moments (or torque) can be measured \cite{couetard-93}. Recently, an application consisting in six components measurements of the actions torsor in cutting process was carried out for the case of high speed milling \cite{couetard-A-darnis-01}, drilling \cite{laporte-AA-darnis-07}, \cite{yaldiz-AA-isik-07}, etc. Cahuc et al., in \cite{cahuc-AA-battaglia-01} presents another use of this six components dynamometer in an experimental study: the taking into account of the cut moments allows a better machine tool power consumption evaluation. The present paper is dedicated to the six components dynamometer use to reach with fine accuracy a vibrations dynamic influence evaluation in turning on the system Workpiece/Tool/Machine tool (\textbf{WTM}). 

The concepts of slide block system, moments and torsor (forces and moments) are directly related to the work, undertaken to the previous century beginning, on the mathematical tool "torsor" \cite{stawell-00}. Unfortunately, until now the results on the cutting forces are almost still validated using platforms of forces (dynamometers) measuring the three components of those \cite{lian-A-huang-07}, \cite{marui-A-kato-83b}. The actions torsor is thus often truncated because the torsor moment part is probably neglected fault of access to an adapted metrology \cite{yaldiz-unsacar-06b}, \cite{yaldiz-unsacar-06a}.

However, recent experimental studies showed the existence of cutting moments to the tool tip \cite{cahuc-gerard-06}, \cite{laheurte-AA-battaglia-03}, \cite{laheurte-A-cahuc-02}, \cite{toulouse-AA-gerard-97a}. Their taking into account allows thus the best machine tools output approaches \cite{cahuc-AA-battaglia-01}, \cite{darnis-A-couetard-00}. Nowadays, the use of a dynamometer measuring the mechanical actions torsor six components \cite{bisu-AAAAA-ispas-07a}, \cite{couetard-00a}, \cite{laporte-AA-darnis-07} allows a better cut approach and should enable to reach new system \textbf{WTM} vibrations properties in the dynamic case.

Moreover, the tool torsor has the advantage of being transportable in any space point and in particular at the tool tip in O point. The study which follows about the cut torsor carries out in several stages including two major; the first relates to the forces analysis and second is dedicated to a first moments analysis to the tool tip during the cut.

After this general information on the torsor concept, an analysis of the efforts (section~\ref{sec:2}) exerted in the cutting action is carried out. Thanks to this analysis and experimental tests, we establish that the points of variable application exerted on the tool type are included in a plane, more precisely following an ellipse. Then, accordingly in section~\ref{PremieresAnalDesMoments}, the first moments analysis obtained at the tool tip in O point is performed. The torsor central axis is required (section~\ref{AxeCentral}) and the central axes beams deduced from the multiple tests strongly confirm the moments presence to the  tool tip.

Some chip characteristics are presented in the section~\ref{GEomEtriePiEceEtCopeau}, and compaired to experimental data base. Before concluding, a correlation (section~\ref{CorrElationDEplacementsForces}) displacements / cutting forces shows that these two dual elements evolve in a similar way following ellipses having the similar properties.

\section{Forces analysis}
\label{sec:2}

\subsection{Tests results}
\label{ResultEssai}

\begin{figure}[htbp]
	\centering
		\includegraphics[width=0.48\textwidth]{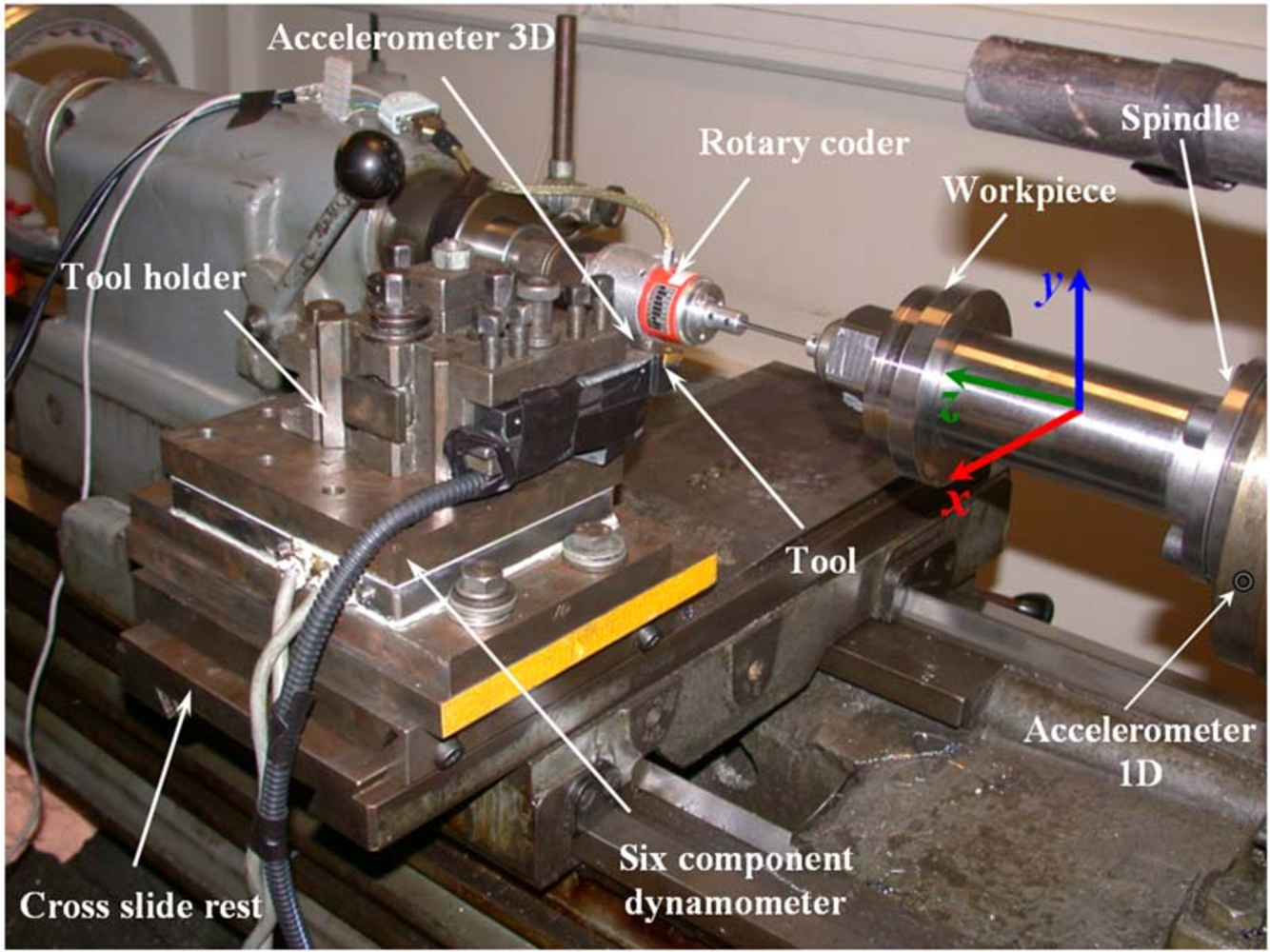}
	\caption{Experimental device and associated measurement elements.}
	\label{fig1}
\end{figure}

The experiments are performed within a framework similar to that of Cahuc et al \cite{cahuc-gerard-06}. For each test, except the feed rate values, all the turning parameters are constant. The mechanical actions are measured according to the feed rate (f) using the six components dynamometer \cite{couetard-93} following the method initiated in Toulouse \cite{toulouse-98}, developed and finalized by Cou\'etard \cite{couetard-00a} and used in several occasions \cite{cahuc-AA-battaglia-01}, \cite{couetard-A-darnis-01}, \cite{darnis-A-couetard-00}, \cite{laheurte-AA-battaglia-03}, \cite{laheurte-A-cahuc-02}. On the experimental device (Fig.~\ref{fig1}) the instantaneous spindle speed is permanently controlled (with an accuracy of 1\%) by a rotary encoder directly coupled with the workpiece. During the tests the insert tool used is type TNMA~16~04~12 carbide not covered, without chip breeze. The machined material is an alloy of chrome molybdenum type 42CrMo24. The test-workpieces are cylindrical with a diameter of 120~mm and a length of 30~mm. They were designed starting from the Finite Elements Method being coupled to a procedure of optimization described in \cite{bisu-AA-k'nevez-07}. 

Moreover, the tool geometry is characterized by the cutting angle $\gamma$, the clearance angle $\alpha$, the inclination angle of edge $\gamma_{s}$, the direct angle $\kappa_{r}$, the cutting edge radius $r_{\epsilon}$, and the sharpness radius R \cite{laheurte-04}. In order to limit to the wear appearance maximum along the cutting face, the tool insert is examined after each test and is changed if necessary (Vb $\leq$ 0.2 mm ISO 3685). The tool parameters are detailed in the Table \ref{tabl-1}.

\begin{table}[htbp]
	\centering
		\begin{tabular}{|c|c|c|c|c|c|}
\hline
 $\gamma$ & $\alpha$ & $\lambda_{s}$ & $\kappa_{r}$ & $r_{\epsilon}$ & R\\
\hline
-6$^{\circ}$ & 6$^{\circ}$ & -6$^{\circ}$ & 91$^{\circ}$ & 1,2 mm & 0,02 mm\\
\hline
\end{tabular}
\caption{Tool geometrical characteristics.}
	\label{tabl-1}
\end{table}

\begin{figure}[htbp]
	\centering
		\includegraphics[width=0.48\textwidth]{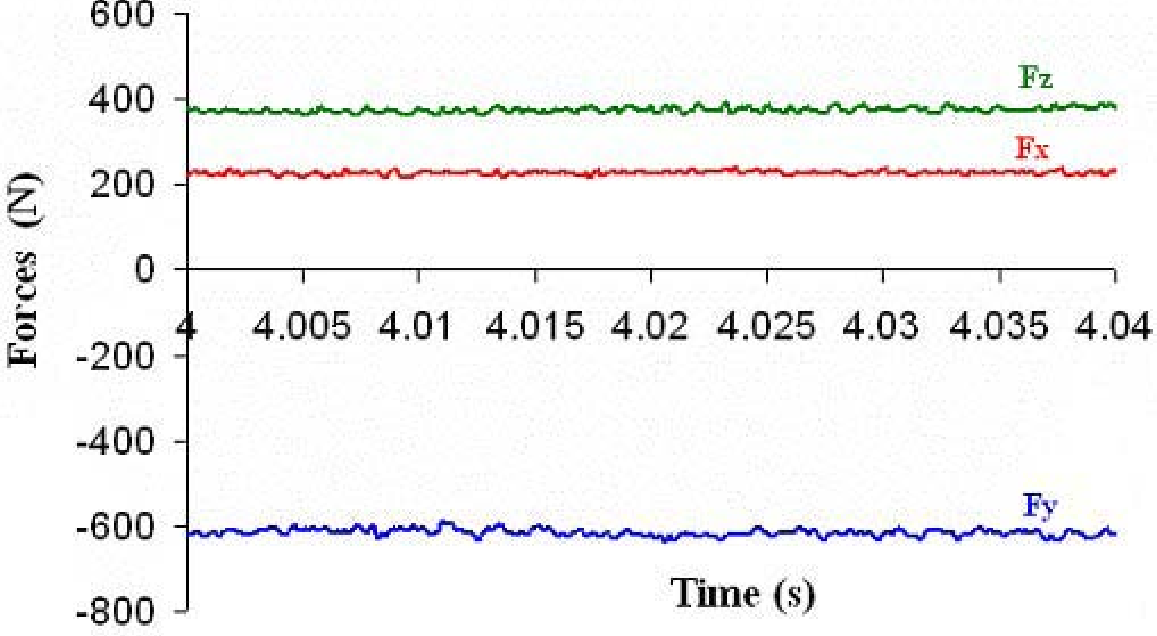}
	\caption{Signals related to the resultant components of cutting forces following the three cutting directions; test case using parameters ap = 2~mm, f = 0,1~mm/rev and N = 690~rpm.}
	\label{Fig-2}
\end{figure}

\begin{figure}[htbp]
	\centering
	\includegraphics[width=0.48\textwidth]{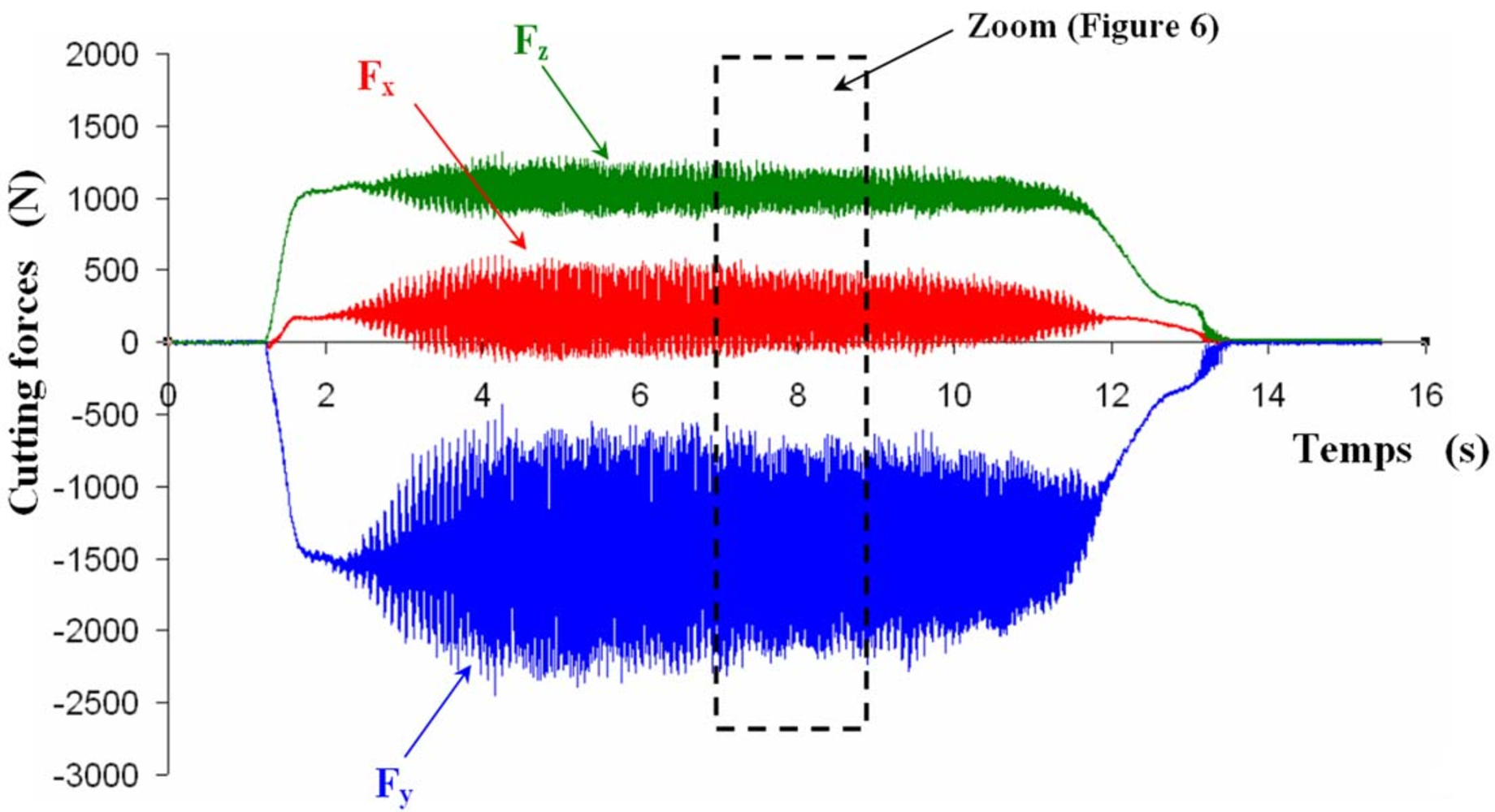}
	\caption{Signals related to the resultant components of cutting forces following the three (x, y, z ) cutting directions; test case using parameters ap = 5~mm, f = 0,1~mm/rev and N = 690~rpm.}
	\label{Fig-3}
\end{figure}

Two examples of resultant efforts measurements applied to the tool tip are presented: one of these for the stable case, ap =~2~mm (Fig.~\ref{Fig-2}), and other for the case with instability, ap =~5~mm (Fig.~\ref{Fig-3}). In the stable case it appears that the force components amplitudes remain almost independant from time parameter. Thus, the amplitude variation is limited to 1 or 2~N around their nominal values, starting with 200~N for ($F_{x}$) and until 600~N for ($F_{y}$). These variations are quite negligible. Indeed the nominal stress reached, the component noticed as the lowest value is the ($F_{x}$) one, while the highest in the absolute value is (the $F_{z}$) one. While taking as reference the absolute value of ($F_{x}$) the following relation between these three components comes:
 $\left|F_{x}\right|$ = $\frac{\left|F_{z}\right|}{2}$ = $\frac{\left|F_{y}\right|}{3}$. 
 
In the unstable case we observe that the efforts components on the cutting axis ($F_{y}$) has the most important average amplitude (1~500~N). It is also the most disturbed ($\pm$~700~N) with oscillations between -2~200~N and -800~N. In the same way the effort according to the feed rate axis ($F_{z}$) has important average amplitude (1~000~N) and the oscillations have less width in absolute value ($\pm$~200~N) but also in relative value ($\pm$~20\%). As for the effort on the radial direction ($F_{x}$) it is weakest on average (200~N) but also most disturbed in relative value ($\pm$~200~N). These important oscillations are the tangible consequence of the contact tool/workpiece frequent ruptures and thus demonstrate the vibration and dynamical behaviour of the system \textbf{WTM}.

Finally, we note that the amplitudes of all these efforts components applied to the tool tip are slightly decreasing functions of time in particular for the component according to the cutting axis.
  
\subsection{Frequency analysis}
\label{Anafreq}

\begin{figure}[htbp]
	\centering
		\includegraphics[width=0.48\textwidth]{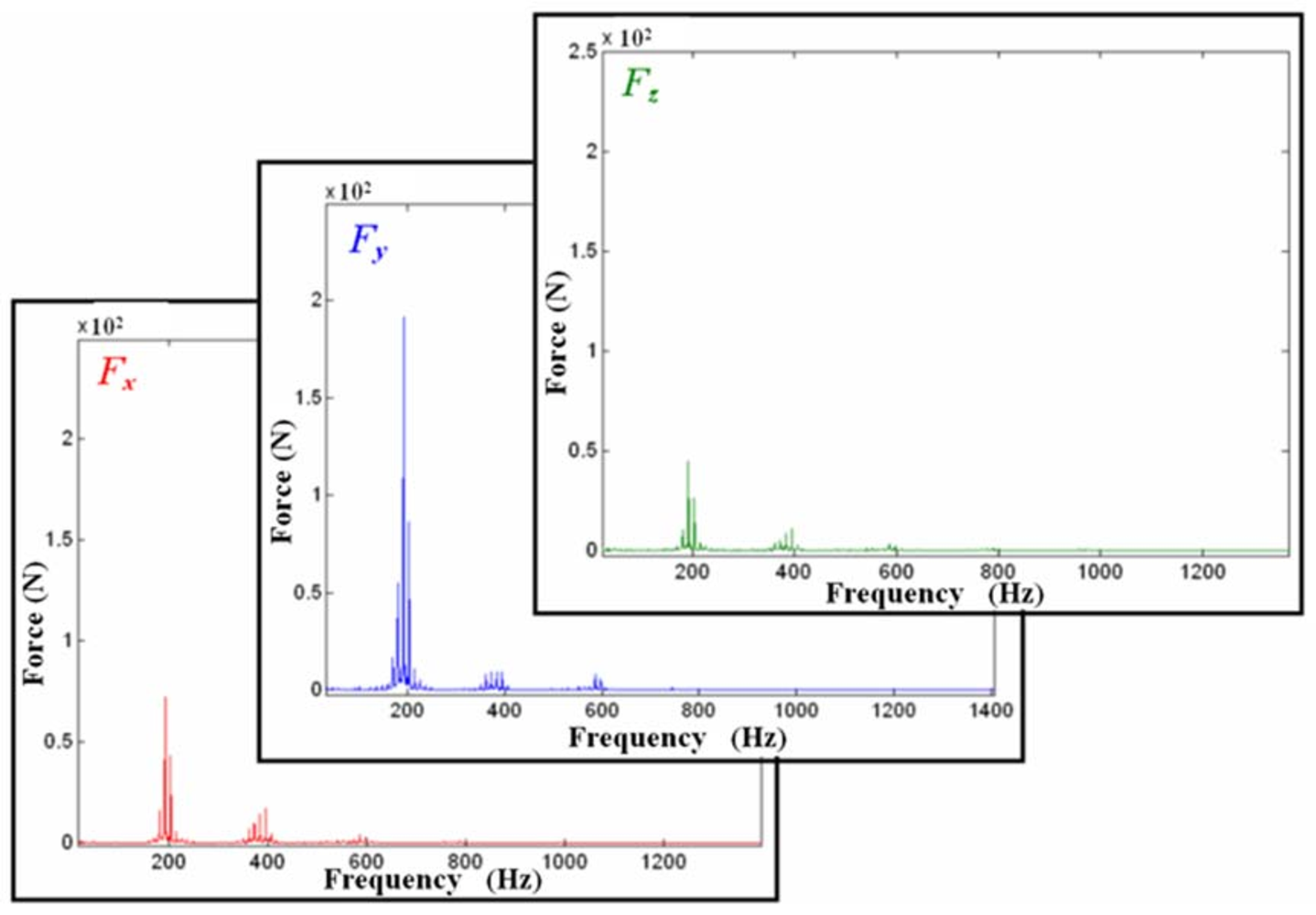}
	\caption{Cutting forces magnitude FFT signal on the three (x, y, z) directions; test case using cutting parameters ap = 5~mm, f = 0,1~mm/rev and N = 690~rpm.}
	\label{Fig-4}
\end{figure}

The signals frequency analysis performed by using FFT function enables to note in Fig.~\ref{Fig-4} the presence of frequencies peaks around 190~Hz. Around this frequency peak, we note for the three forces components, a quite high concentration of energy in a wide bandwidth around 70 Hz (36\% of the fundamental frequency). All things considered, this width of frequency is of the same order of magnitude as observed (13\% of fundamental) by Dimla \cite{dimla-04} for a depth of cut three times lower (ap = 1.5 mm) but for an identical feed rate (f=0.1 mm/rev) and a cutting speed similar.This remark confirms that the efforts components is proportionnal to the depth of cut ap as indicated by Benardos et al., \cite{benardos-A-vosniakos-06}.

Let us recall that the same frequency of 190~Hz was observed in the tool tip displacements case (in conformity with \cite{bisu-AAAAA-ispas-07a}). Consequently, the cutting forces components variations and the self-excited vibrations are influenced mutually, in agreement with \cite{ispas-AA-anghel-99}, \cite{kudinov-70}, \cite{marot-80}, \cite{tansel-A-keramidas-92}. Also, in agreement with research on the dynamic cutting process \cite{moraru-A-rusu-79}, we note that the self-excited vibrations frequency is different from the workpiece rotational frequency which is located around 220~Hz.

\subsection{Forces decomposition}
\label{DEcompositionDesForces}

\begin{figure}[htbp]
	\centering
		\includegraphics[width=0.48\textwidth]{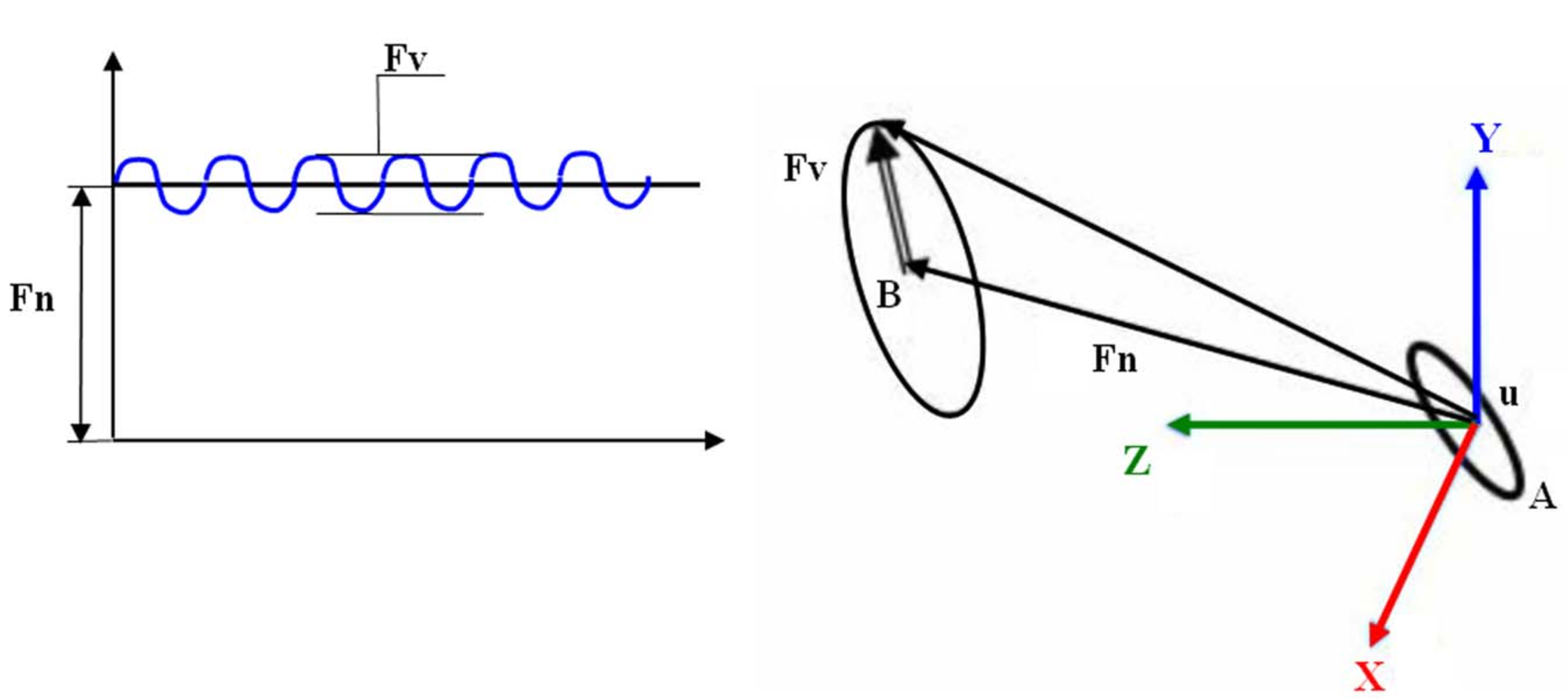}
	\caption{Cutting force $F_{v}$ evolution around the nominal value $F_{n}$; test case using cutting parameters ap = 5~mm, f = 0,1~mm/rev and N = 690~rpm.}
	\label{Fig-5}
\end{figure}

The forces resultant components detailed analysis highlights a plane in which evolves a variable cutting force $F_{v}$ around a nominal value $F_{n}$ (see further). This variable force is an oscillating action (Fig.~\ref{Fig-5}) which generates $u$ tool tip displacements and maintains the vibrations of elastic system block-tool \textbf{BT} \cite{bisu-AA-knevez-06}.

\begin{figure}
	\centering
		\includegraphics[width=0.48\textwidth]{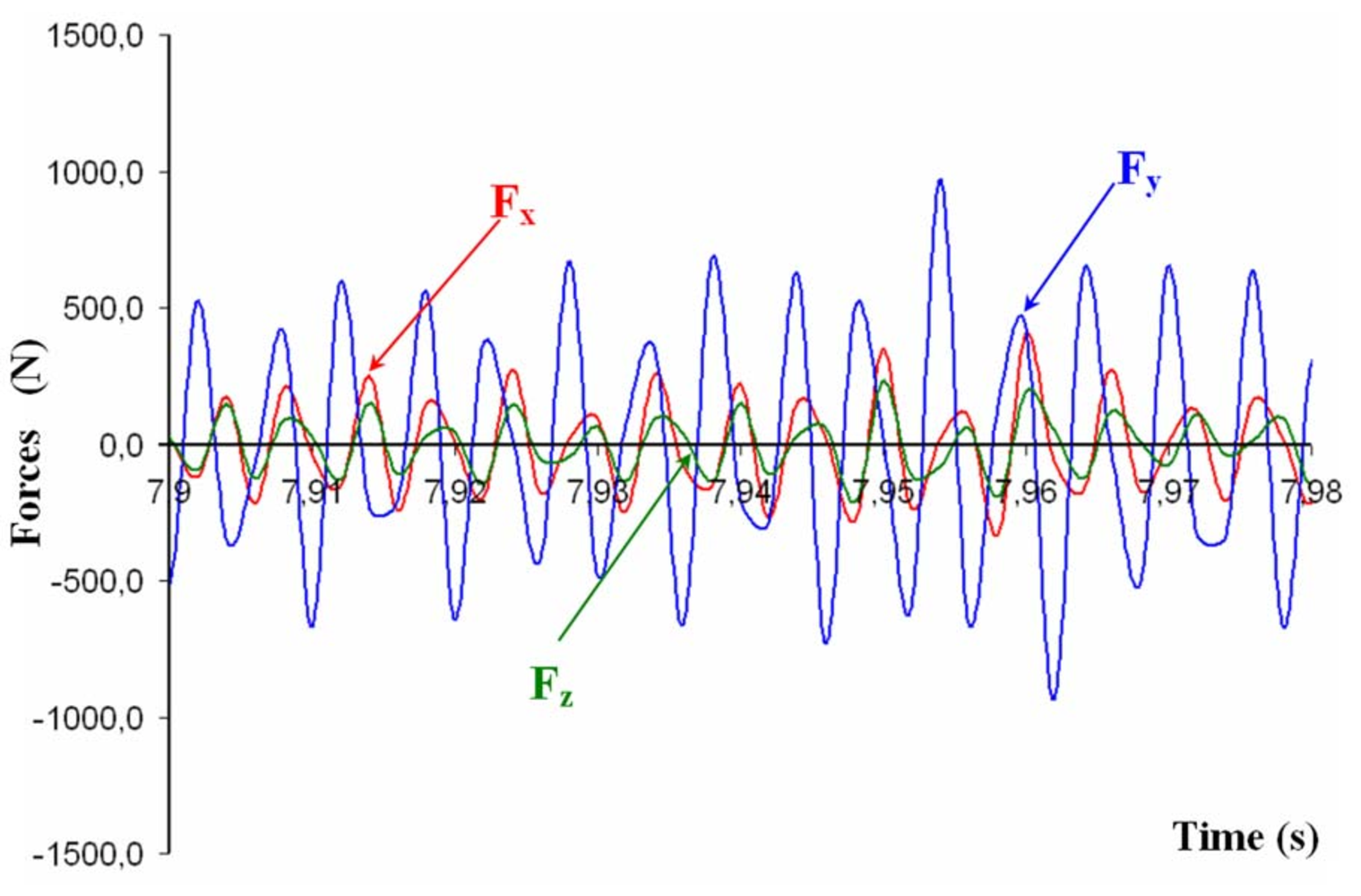}
	\caption{The resultant cutting force components variable zoom on the three cutting directions; test case using parameters ap = 5~mm, f = 0,1~mm/rev and N = 690~rpm.}
	\label{Fig-6}
\end{figure}
      
Thus, the cutting force variable (Fig.~\ref{Fig-5}) and the self-excited vibrations of elastic system \textbf{WTM} are interactive, in agreement with research work \cite{koenigsberger-tlusty-70}, \cite{moraru-A-rusu-79}.

\begin{figure}[h!tbp]
	\centering
		\includegraphics[width=0.48\textwidth]{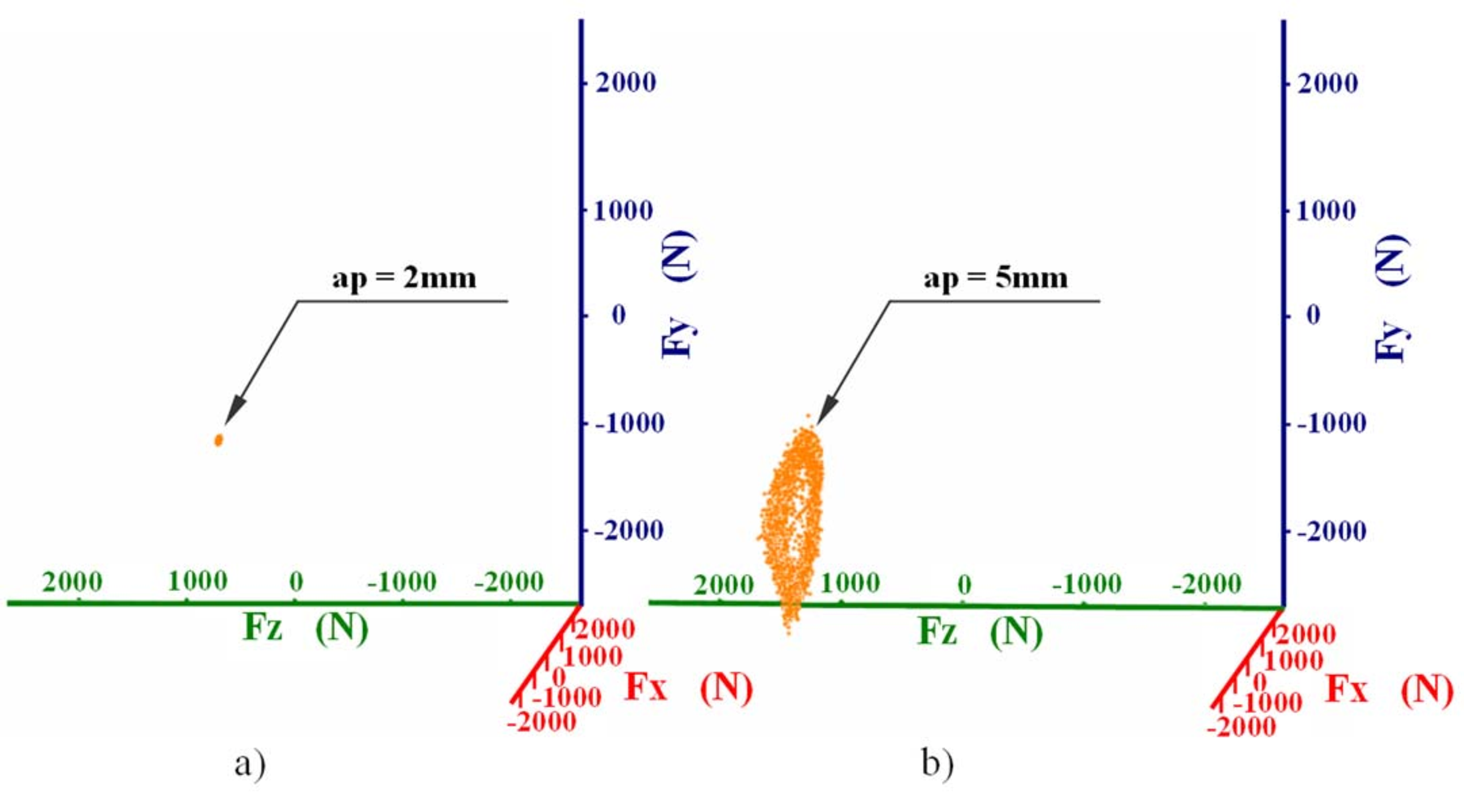}
	\caption{Stable process (a) and unstable process (b).}
	\label{Fig-7}
\end{figure}

The cutting forces variable part can be observed and compared. Not to weigh down this part, the cutting forces analysis is voluntarily below restricted at only two different situations:

\begin{itemize}
\item[-] stable process using cutting depth ap = 2~mm (Fig.~\ref{Fig-7}a),
\item[-] unstable process (with vibrations) using cutting depth ap = 5~mm (Fig.~\ref{Fig-7}b)

\end{itemize}

The vibrations effects on the variable forces evolution are detailed on the Fig.~\ref{Fig-7}b. Moreover, the variable forces and displacements analysis associated with the tool tip at the time of the unstable process shows that the forces variation ratio is equivalent to the tool tip displacements variation ratio (in conformity with \cite{bisu-07}). This aspect will be quantified further. Also we concentrate our attention mainly on the unstable case (ap =~5~mm).

\subsection{Plane determination attached of the forces application points}
\label{DEcompositionPlanDesForces}

\begin{figure}[htbp]
	\centering		\includegraphics[width=0.48\textwidth]{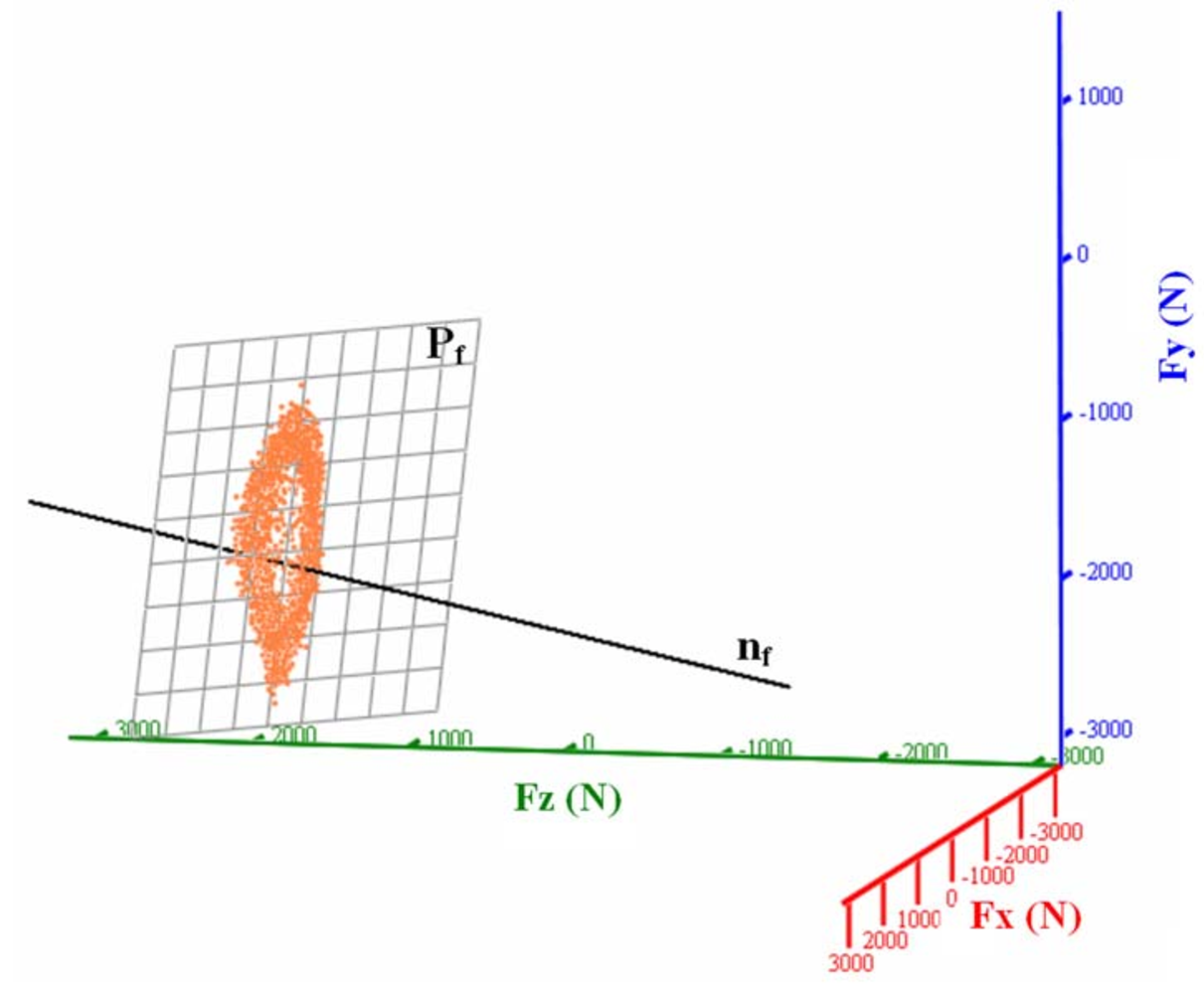}
	\caption{The ellipse plane $P_{f}$ attached of range of the forces application points considering ap = 5~mm, f = 0,1~mm/rev and N = 690~rpm.}
	\label{Fig-8}
\end{figure}

The tests analysis shows that, in the vibratory mode, the load application points describe an ellipse (Fig.~\ref{Fig-8}), that is not the case in the stable mode (without vibrations, Fig.~\ref{Fig-7}a)).

The method used in \cite{bisu-AA-knevez-08} to determine the tool tip displacements plane is taken again here (in conformity with Appendix, section \ref{sec:DEterminationDuPlanLieuDesPointsDApplicationDesEffortsSurLOutil}) to establish the plane $P_{f}$, place of the load application points, characterized by its normal $\vec{n_{f}}$ (Table~\ref{tabl-2}).

\begin{table}[h!tbp]
	\centering
	\begin{tabular}{|c|c|c|c|c|}
\hline
\multicolumn{2}{|c|}{ap = 5 mm}& x & y & z \\

\hline
\multicolumn{2}{|c|}{Normal} &\multicolumn{3}{|c|}{$\vec{n_{f}}$}\\
\hline
 \multirow{4}* &{0.1} & 0.46 & - 0.1 & - 0.882\\
{f (mm/rev)} & {0.075} & 0.419& - 0.097 & - 0.903\\
&0.0625 & 0.292 & - 0.113& - 0.95  \\
& 0.05 & 0.245 & - 0.107 & - 0.964\\
\hline

\end{tabular}
\caption{Directory normal of the plane $P_{f}$ attached of forces application points on the tool; case study using ap = 5~mm, f = 0,1~mm/rev and N = 690~rpm.}
	\label{tabl-2}
\end{table}

\begin{figure}[htbp]
	\centering
		\includegraphics[width=0.48\textwidth]{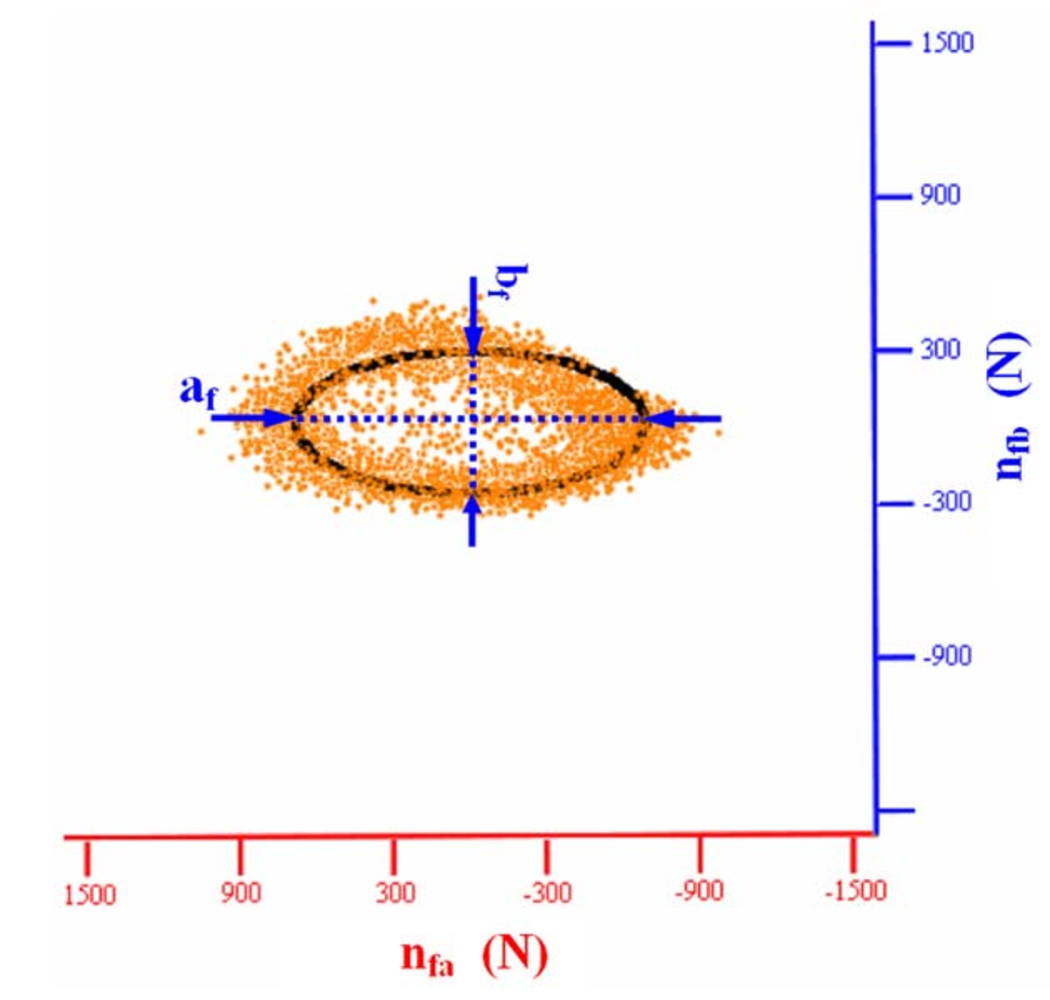}
	\caption{Ellipse approximation attached of forces application points; case study using parameters ap = 5~mm, f = 0,1~mm/rev and N = 690~rpm.}
	\label{Fig-9}
\end{figure}

As for the tool tip displacements study related, a new reference system $(\vec{n_{fa}}, \vec{n_{fb}})$ is associated at the load application points ellipse place. In this new reference system (in conformity with section \ref{sec:ApproximationDeLEllipse}) the ellipse large (respectively small) axis dimensions $a_{f}$ (respectively $b_{f}$) are obtained (Fig.~\ref{Fig-9}). The values of $a_{f}$, $b_{f}$ as that of their ratio $a_{f}$ / $b_{f}$ are consigned in the Table~\ref{tabl-3}. It should be noted that $a_{f}$ and $b_{f}$ are increasing feed rate $(f)$ functions. It is thus the same for the ellipse surface which grows with $(f)$, in perfect agreement with the mechanical power injected into the system which is also an increasing feed rate function. On the other hand, the ratio $a_{f}$ / $b_{f}$ is a decreasing feed rate function. Thus the ellipse elongation evolves in a coherent way in dependence with the feed rate.

\begin{table}[htbp]
	\centering
		\begin{tabular}{|c|c|c|c|}
\hline
 $f$ (mm/rev) & $a_{f}$ (N) & $b_{f}$ (N) & $a_{f}$/$b_{f}$ \\
\hline
 $0.1$ & $690$ & $280$ & $2.46$\\
\hline
$0.075$ & $545$ & $220$ & $2.47$ \\
\hline
$0.0625$ & $365$ & $147$ & $2.48$ \\
\hline
$0.05$ & $220$ & $88$ & $2.5$ \\
\hline
\end{tabular}
\caption{ Large and small ellipse axes attached of forces application points depending on feed rate parameter; case study using ap = 5~mm, f = 0.1~mm/rev and N = 690~rpm.}
	\label{tabl-3}
\end{table}

Let us look at now the moments evolution to the tool tip.

\section{First moment analysis}
\label{PremieresAnalDesMoments}

\subsection{Experimental results}
\label{Experimental results}

\begin{figure}[htbp]
	\centering
		\includegraphics[width=0.48\textwidth]{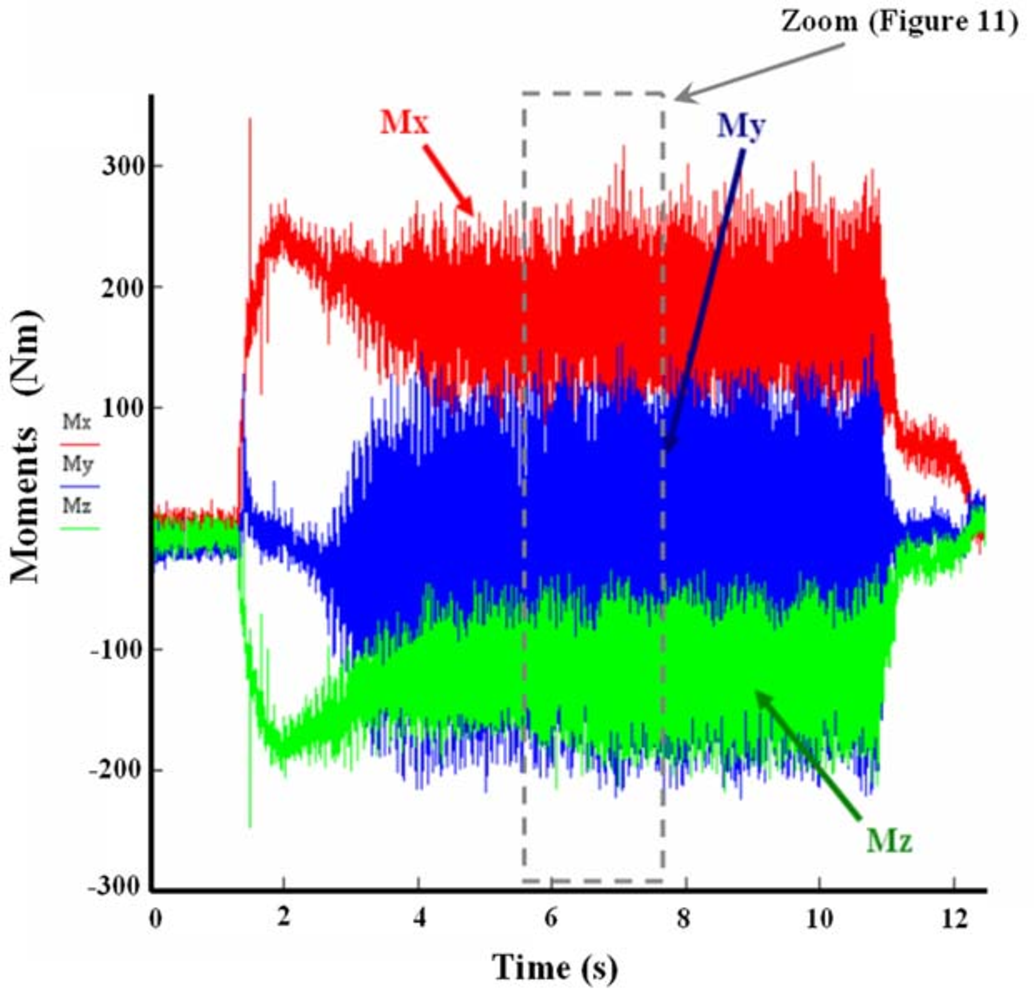}
	\caption{The moments components time signals that action on the tool tip following the three directions; case study using ap = 5~mm, f = 0.1~mm/rev and N = 690~rpm.}
	\label{Fig-10}
\end{figure}

\begin{figure}[htbp]
	\centering
		\includegraphics[width=0.48\textwidth]{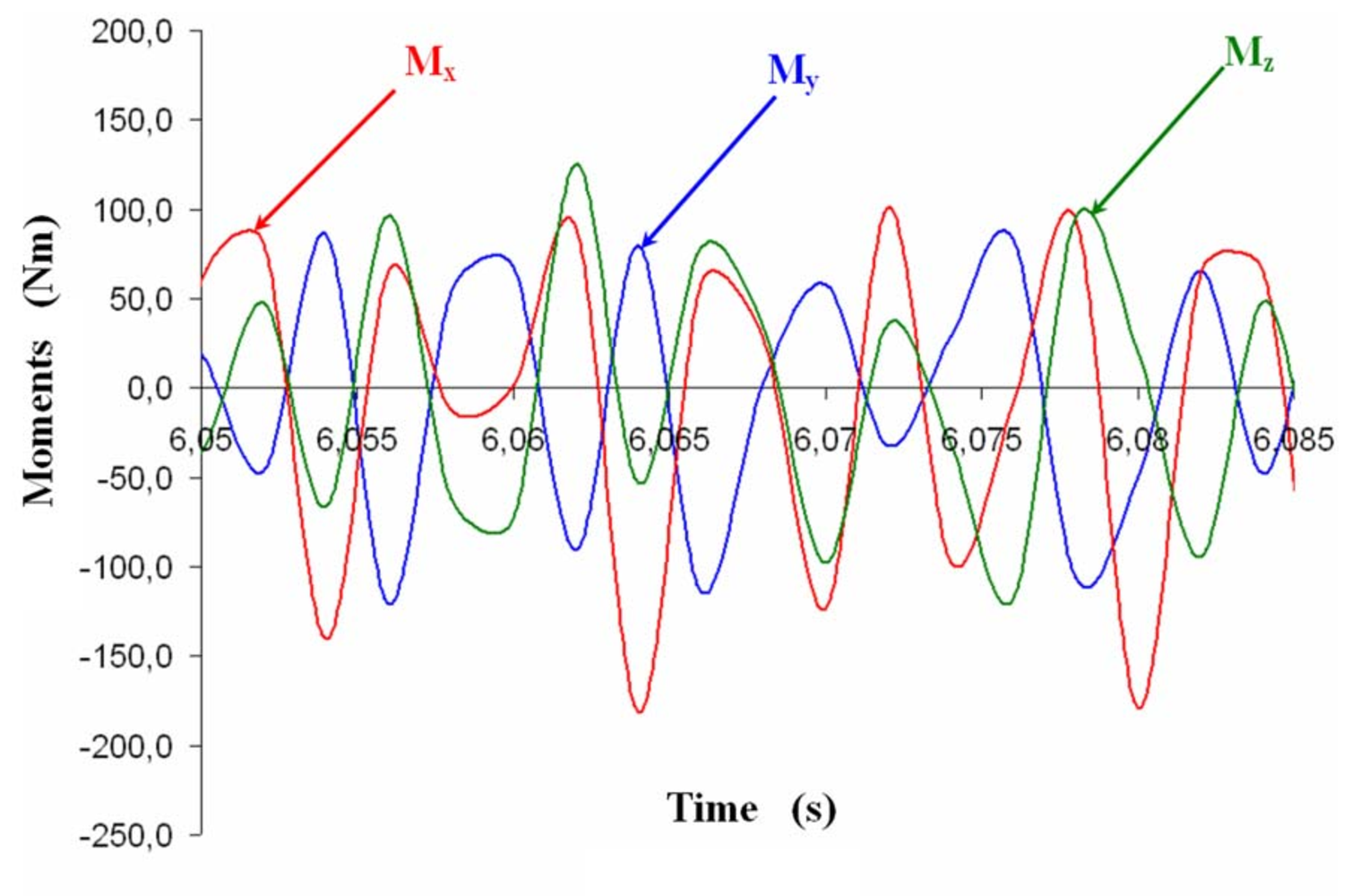}
	\caption{The moments components variable part zoom that action on the tool tip following the three cutting directions; case study ap = 5~mm, f = 0,1~mm/rev and N = 690~rpm.}
	\label{Fig-11}
\end{figure}

For each test, the mechanical actions complete torsor is measured according to the method already detailed in the section~\ref{ResultEssai} \cite{couetard-00a}. Measurements are taken in the six components dynamometer transducer O' center and then transported to the  tool point O via the moment transport traditional relations \cite{brousse-73}. As for the forces, the moments variable part is extracted from measurements. An example of the results measurement is given on the Fig.~\ref{Fig-10} which zooms on the moments variable part is presented in the Fig.~\ref{Fig-11}.

Taking into account the recordings chaotic aspect obtained, an accurate moments components frequency analysis is necessary. It is the object of the following section.

\subsection{Moments frequency analysis}
\label{sec:AnalyseFrEquentielleDesMoments}

\begin{figure}[htbp]
	\centering
		\includegraphics[width=0.48\textwidth]{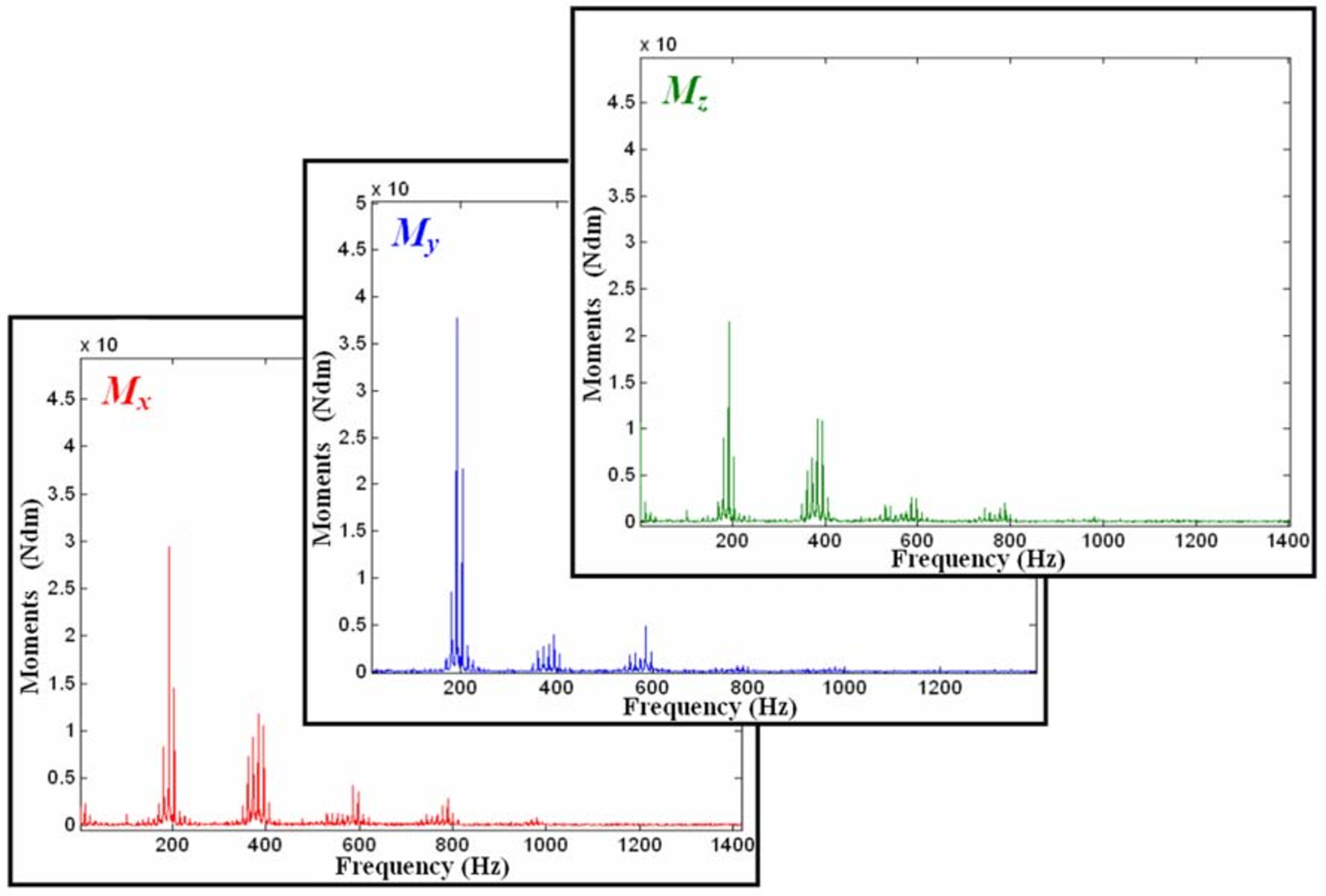}
	\caption{Signal FFT related to the moment on the three (x, y, z) directions; case study using ap = 5~mm, f = 0.1~mm/rev and N = 690~rpm.}
	\label{Fig-12}
\end{figure}

An example of moments signals frequency analysis during the vibratory cutting is presented in the Fig.~\ref{Fig-12}. As for the forces analysis, the moments components FFT shows that the most important frequency peak is localized around 190~Hz. 

Moreover of all the components, the most important fundamental amplitude is that corresponding to the moments components following to the cutting axis $(y)$ as its transport at the tool tip confirms it. It should be noted that the force component following to this same axis is also most important but has obviously no influence on the moments related to this axis due to co-linearity of these two elements. Conversely, the least important vibration amplitude is that of the moment component located on z axis. However, the number of revolutions being important according to this axis the $M_{z}$ component contribution to the torques power consumption remains significant.

\begin{figure}[htbp]
	\centering
		\includegraphics[width=0.46\textwidth]{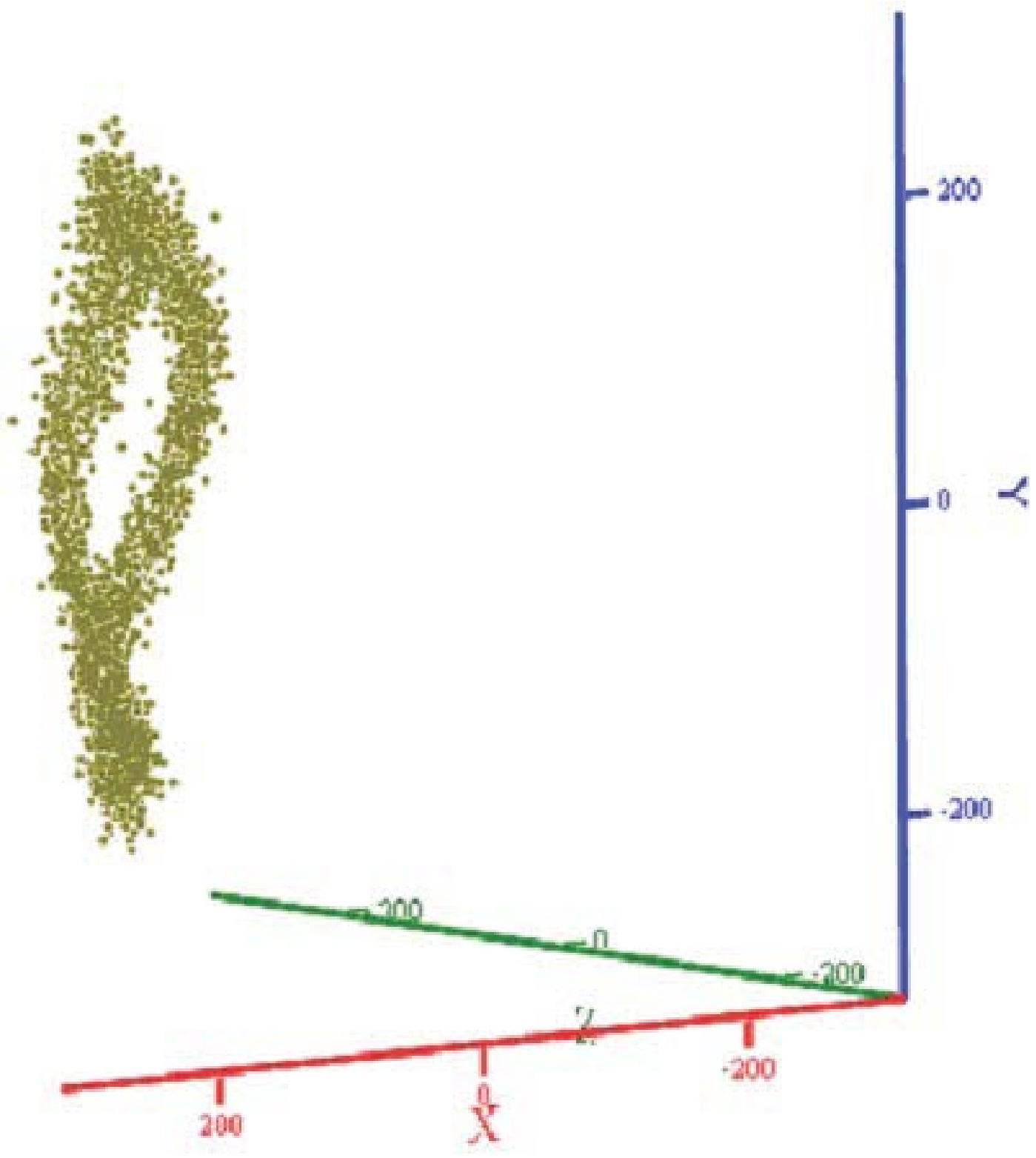}
	\caption{Moments place space representation.}
	\label{Fig-13}
\end{figure}

The appearance of other peaks, which are harmonics, slightly modifies the three-dimensional moments representation (Fig.~\ref{Fig-13}) which is not exclusively any more in a plane although the ellipse essence is in a plane. This representation approaches a light form of letter eight contrary to the elliptic planar form efforts characteristic.

\section{Central axis}
\label{AxeCentral}

\subsection{Central axis determination}
\label{DEterminationDeLAxeCentral}

It is well-known that with any torsor, it is possible to associate a central axis (except the torsor of pure moment), which is the single object calculated starting from the torsor six components \cite{brousse-73}. 

A torsor $\left[A\right]_{O}$ in a point O is composed of a forces resultant $\vec{R}$ and the resulting moment $\vec{M_{O}}$ :

\begin{eqnarray}
	[A]_{O} = \left\{
	\begin{array}{c}
	\vec{R} ,\\
  \vec{M_{O} .}
\end{array}\right.
\end{eqnarray}

\begin{figure}[htbp]
	\centering
		\includegraphics[width=0.48\textwidth]{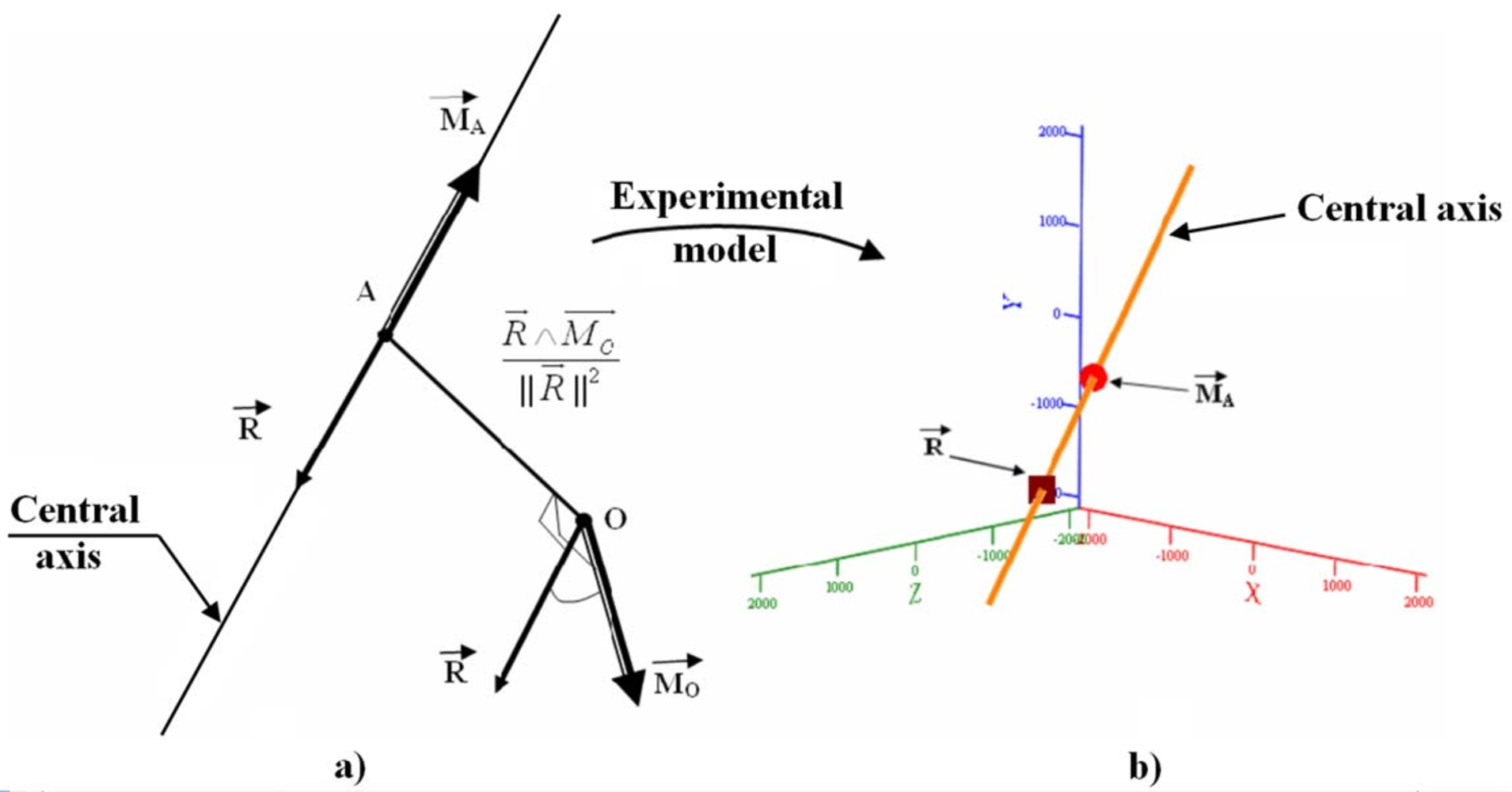}
	\caption{Central axis representation (a) and of the co-linearity between vector sum $\vec{R}$ and minimum moment  $\vec{M_{A}}$ on central axis (b).}
	\label{Fig-14}
\end{figure}

The central axis is the line defined classically by:

\begin{equation}
	\vec{OA}=\frac{\vec{R}\wedge\vec{M}_{O}}{\left|\vec{R}^{2}\right|}+\lambda\vec{R} ,
\end{equation}

where O is the point where the mechanical actions torsor was moved (here the tool tip) and A the current point describing the central axis. \vec{OA} is thus the vector associated with the bi-point [O, A] (Fig.~\ref{Fig-14}).

This line (Fig.~\ref{Fig-14}-(a)) corresponds to points geometric place where the mechanical actions moment torsor is minimal. The central axis calculation consists in determining the points assembly (a line) where the torsor can be expressed according to a slide block (straight line direction) and the pure moment (or torque) \cite{brousse-73}.

The central axis is also the points place where the resultant cutting force is co-linear with the minimum mechanical moment (pure torque). The test results enable to check for each point of measurement the co-linearity between the resultant cutting force $\vec{R}$ and moment $\vec{M_{A}}$ calculated related to the central axis (Fig.~\ref{Fig-14}-(b)). 

\begin{figure}[htbp]
	\centering
		\includegraphics[width=0.48\textwidth]{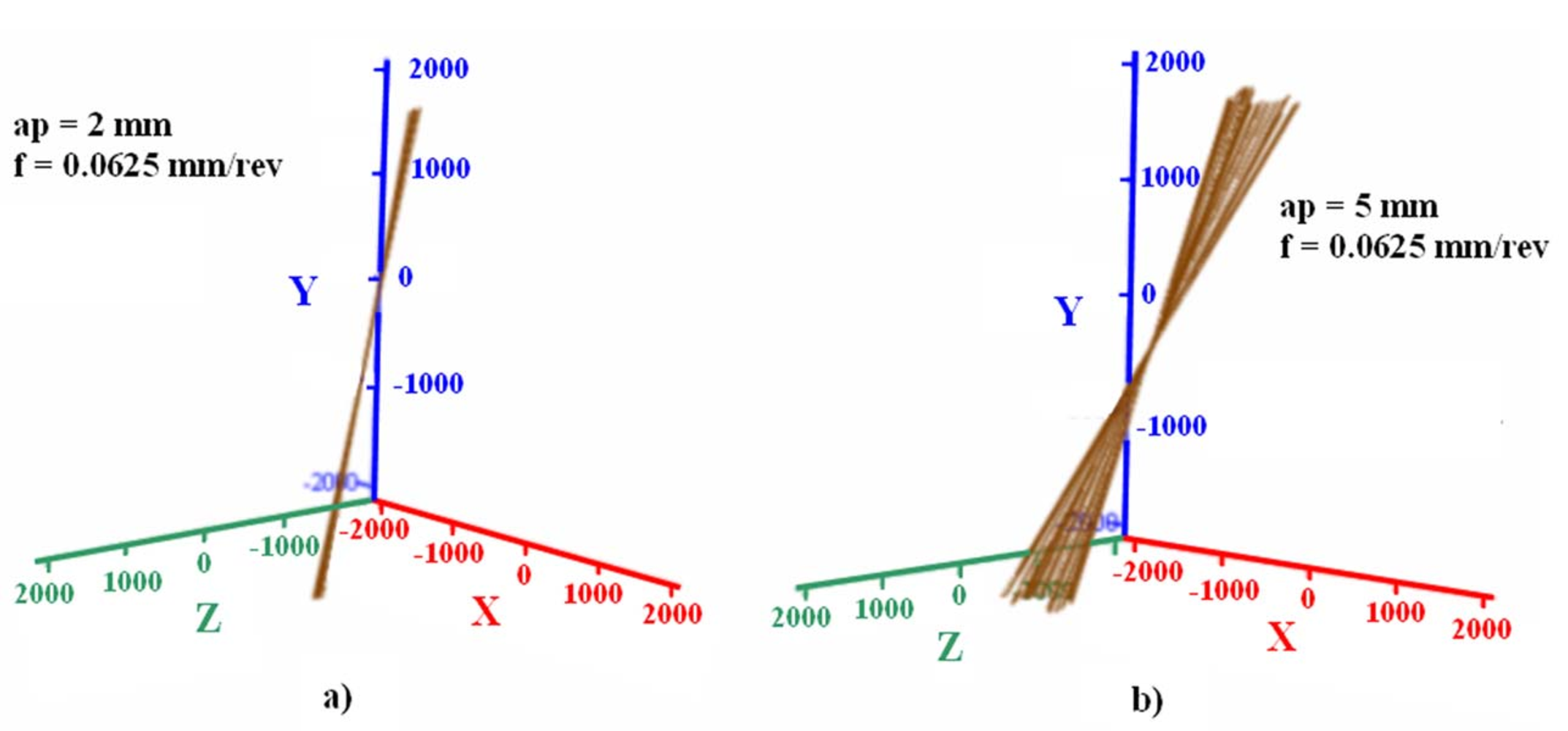}
	\caption{Central axes representation obtained for 68~rpm the workpiece speed and feed rate f = 0,0625~mm/rev; a) - stable process ap = 2~mm; b) - unstable process ap = 5~mm.}
	\label{Fig-15}
\end{figure}

The meticulous examination of the mechanical actions torsor six components shows that the forces and the moment average values are not null. For each measure point, the central axis is calculated, in the stable (Fig.~\ref{Fig-15}-(a)) and unstable mode (Fig.~\ref{Fig-15}-(b)). In any rigour the case ap = 2~mm should be described as quasi-stable movement, because the vibrations exist but their amplitudes are very low - of the order of the $\mu m$ -, thus quasi null compared to the other studied cases. Considering the cutting depth value ap = 3~mm, the recorded amplitude was 10 times more important.   

In the presence of vibrations (ap = 5~mm) for a 68~rpm the workpiece speed (44 points of measure by rpm), the dispersive character of the central axes beam, compared to the stable mode, can be observed, where this same beam is tightened more and less tilted compared to the normal axis on the plane (x,y). This central axes dispersion can be explained by the self-excited vibrations which cause the variable moment generation.

\subsection{Analysis of central axis moments related}
\label{AnalyseDesMomentsALAxeCentral}

\begin{figure}[htbp]
	\centering
		\includegraphics[width=0.48\textwidth]{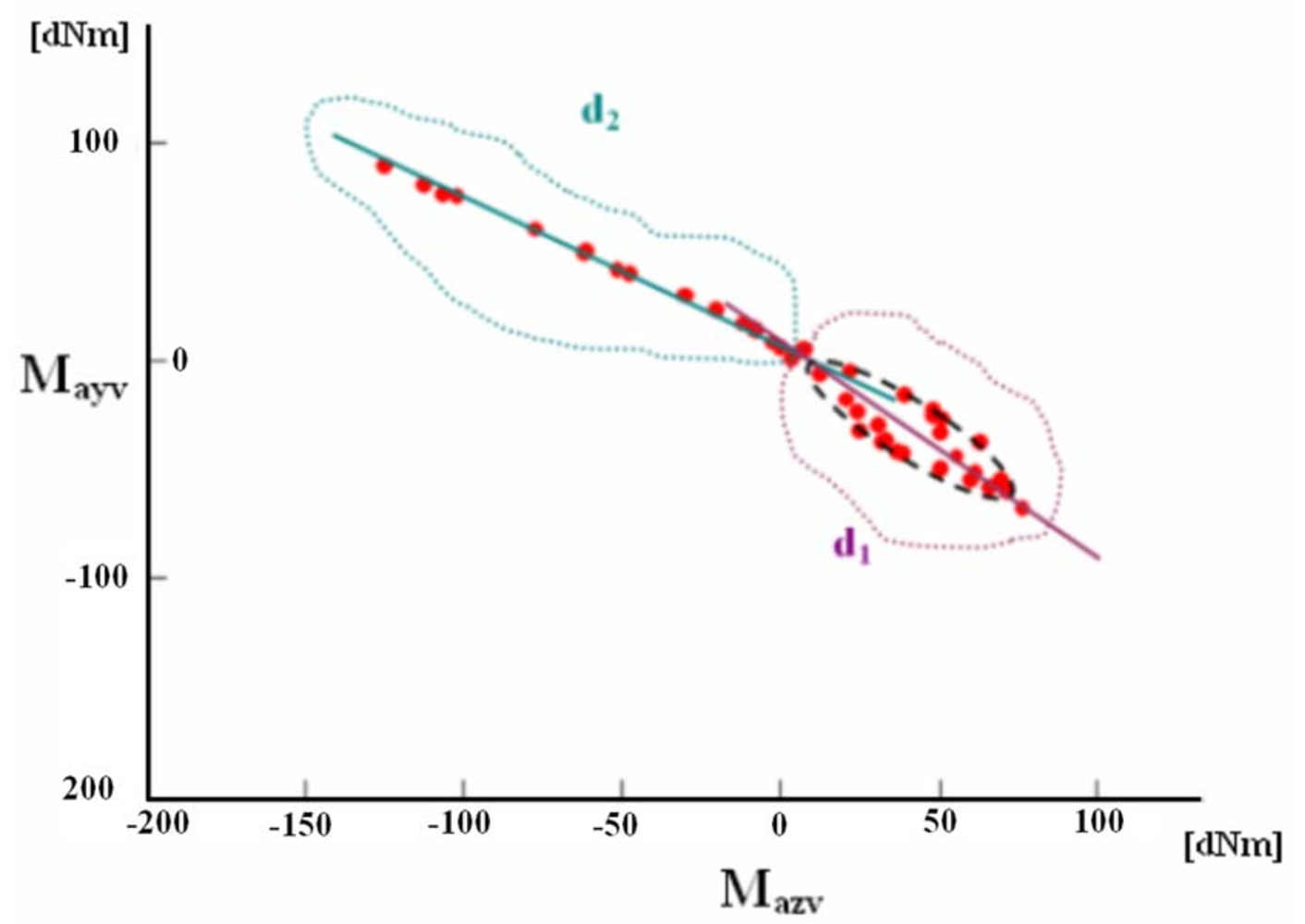}
	\caption{Moment representation related of central axes; case study using ap = 5~mm, f=0,1~mm/rev and N = 690~rpm.}
	\label{Fig-16}
\end{figure}

While transporting the moment from tool tip to the central axis, the minimum moment (pure torque) \vec{M_{A}} is obtained. From the moment values to the central axis, the constant and variable part of this one is deduced. As for the efforts, the variable part is due to the self-excited vibrations as revealed below (Fig.~\ref{Fig-16}).

Using this decomposition, the moments contribution on the zones of contact tool / workpiece / chip is expressed. The observations resulting from the analysis show that the tool vibrations generate rotations, cause variations of contact and thus generate variable moments, confirming the efforts analysis detailed in the section~\ref{ResultEssai}. This representation enables to express the moments along the three axes of the machine tool: swivel moment in the y direction and the two moment of rotation along x and z directions.

Moments components analysis determined at the central axis allows noting a moments localization mainly on two distinct zones. Taking into account this aspect, the variable moments components are divided into two noted parts $d_{1}$ and $d_{2}$ on to the three directions related to the machine. The components of these variable moments on x, y, z, axes are noted respectively ($M_{axv}$), ($M_{ayv}$) and ($M_{azv}$).

In the vibratory case (ap = 5~mm), the first points family is located (Fig.~\ref{Fig-16}) according to a line $d_{2}$ and $d_{1}$ (the large ellipse axis). In the case without vibrations (ap = 2~mm), the two $d_{1}$ and $d_{2}$ families merged to only one family located according to only one line. Thus, the elliptic form appearance around the right-hand side $d_{1}$ seems quite related to the self-excited vibrations (cas ap = 5~mm) \cite{bisu-AA-knevez-06}. In particular, the frequency associated with the $d_{1}$ part is higher than those associated the $d_{2}$ part. Furthermore, these frequencies are related to the frequencies domain found during moments FFT analysis Fig.~\ref{Fig-12}. Finally, on the central axis, the $d_{1}$ families and $d_{2}$  seem to correspond to distinct elements from the generated surface.

\section{Workpiece and chip geometry}
\label{GEomEtriePiEceEtCopeau}

\subsection{Roughness measurements}
\label{MesuresRugosimEtriques}

In the processes of matter per cutting tool removal, it is well-known that the manufactured pieces surface quality is closely connected to the thrust force imposed on the matter \cite{chen-tsao-06a}. In particular, the more the thrust force exerted on the object surface is important, the more the imprint left by the tool on surface is consequent. The surface roughness east thus closely related to the thrust force intensity. The surface roughness is connected confidentially to the thrust force intensity. The self-excited vibrations have an influence on the workpieces quality surface. Now, in the self-excited vibrations case examined here, we have just noticed that the forces and moments are maximum with the resonance frequency which is situated around 190 Hz.

Taking into account the link between roughness and efforts applied to the surface, we should find a maximum of surface roughness around this frequency. In this purpose, we propose to control the profile of a generator length roughness of the cylindrical manufactured piece. Indeed, in every rotation of the machined piece, the tool leaves a imprint force function applied to the generator point observed which corresponds to a given and known moment. Along a given generator, each point of this one is the force image applied to the surface at a known given moment because the rotation speed of the workpiece is known.

The surface roughness examination along a generator should thus necessarily reveal the amplitudes of roughness associated with the efforts periodically applied by the tool in these points. Also the roughness profile FFT presents along a generator of the cylindrical part manufactured should pass, like the efforts, by a maximum around the frequency of 190 Hz. It is indeed what can be observed on the figure~\ref{Fig-16} where the roughness data FFT analysis shows a frequency peak located around 190 Hz (precisely 191,8 Hz), which is coherent with the previous data.

In addition, the surface roughness analysis gives a total roughness value $R_{t}$ = 1,6~$\mu m$.

\begin{figure}[htbp]
	\centering
		\includegraphics[width=0.52\textwidth]{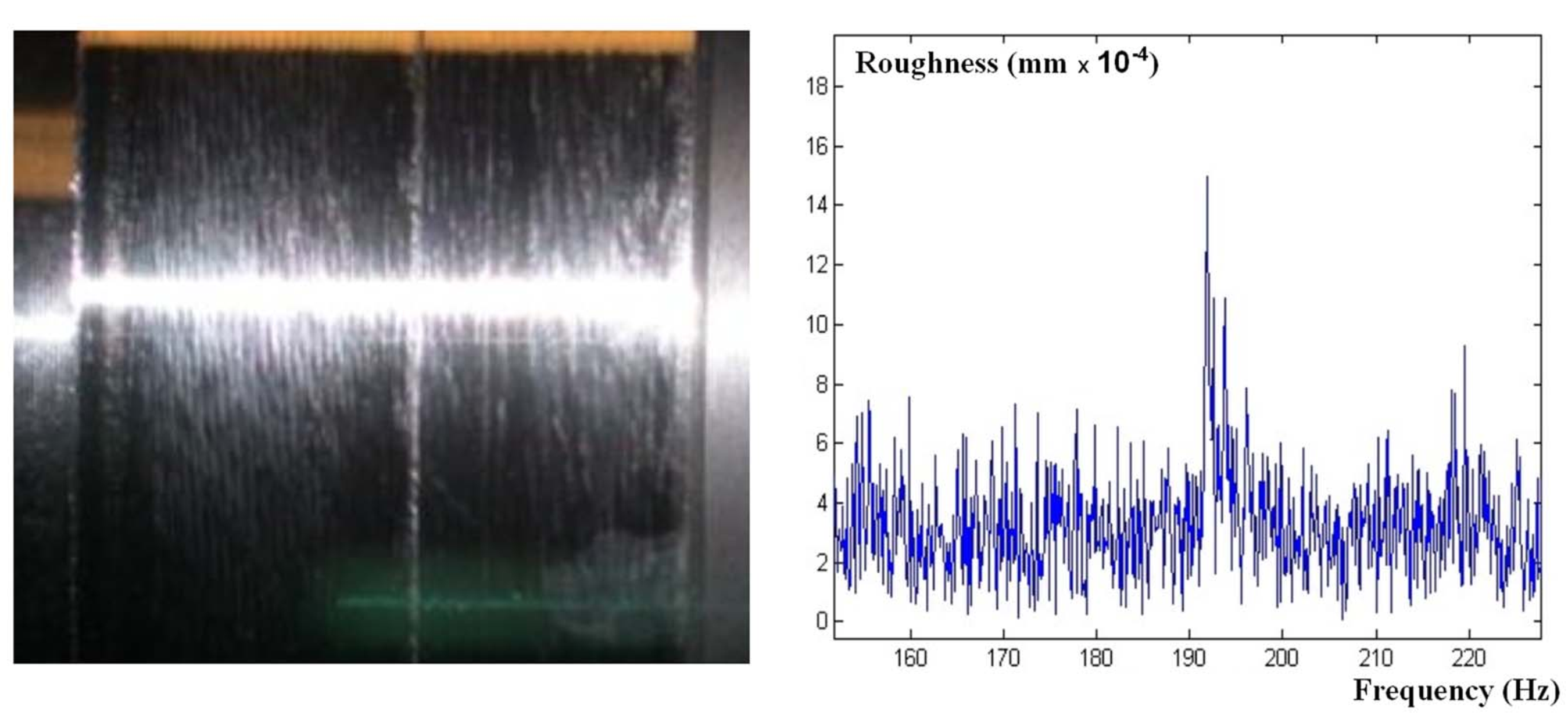}
	\caption{Machined surface and FFT data related to the roughness; case study ap = 5~mm, f = 0,1~mm/rev, N = 690~rpm.}
	\label{Fig-17}
\end{figure}

\subsection{Chip characteristics}
\label{CaractEristiquesDuCopeau}

Chip measurements under the Scanning Electron Microscop were carried out and enabled to determine the thickness variation and the chip width. All chips are type 1.3 (ISO 3685) with undulations.

\begin{figure}[htbp]
	\centering
		\includegraphics[width=0.48\textwidth]{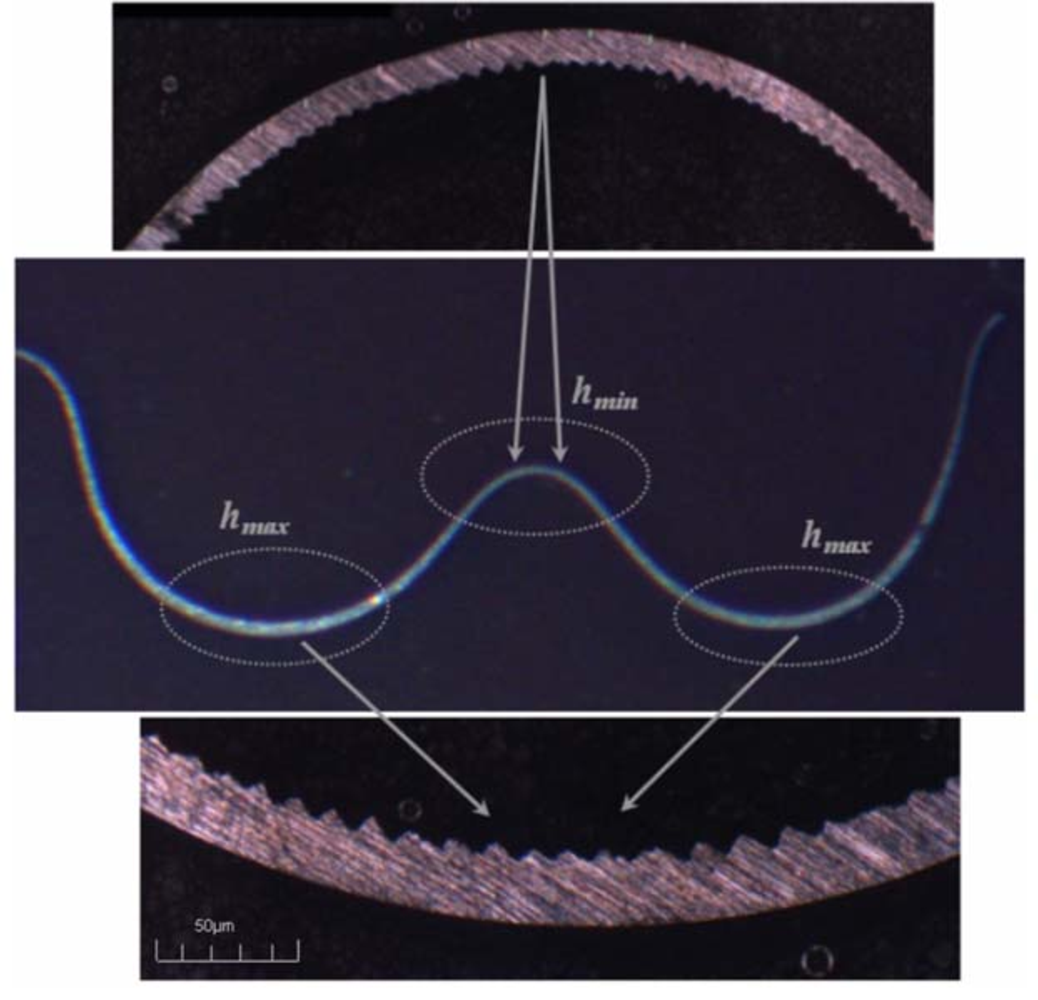}
	\caption{Chip thickness variation evaluation.}
	\label{Fig-18}
\end{figure}

\textit{The chip thickness variations} between the maximum ($h_{max}$) and the minimal ($h_{min}$) thickness are about 2, and feed rate values independent. An example is presented (figure \ref{Fig-18}) for a sample of chip during a test with the feed rate f=0.05~mm/rev. Values obtained $h_{max}$ = 0,23~mm and $h_{min}$ = 0,12~mm.

\textit{The measure of chip length} corresponding to an undulation enables to find the self-excited vibrations frequencies starting from the cutting speed (in conformity with equation 3):

\begin{equation}
	f_{cop} = \frac{V}{l_{0}.\xi_{c}} ,
\end{equation}

with $f_{cop}$ the chip segmentation frequency, $V$ chip speed, $l_{0}$ one chip undulation length and $\xi_{c}$ the chip hardening coefficient.

To determine the total chip length, it is necessary to measure the wavelength, taking into account the rate of hardening phenomenon (in conformity with equation 4) \cite{kudinov-70} of the chip during cutting process and the primary shear angle $\phi$,

\begin{equation}
	\xi_{c} = \frac{\cos(\phi - \gamma)}{\sin (\phi)} .
\end{equation}

In our case, it is measured on the chip undulation length $l_{0}$ = 11~mm, with a chip hardening rate $\xi_{c}$ = 1.8 and the cutting speed $V$ = 238~m/min. A frequency of 206~Hz is then obtained, very near to the frequencies of tool tip displacements or of the load application points during cutting process.

\begin{figure}
	\centering
		\includegraphics[width=0.48\textwidth]{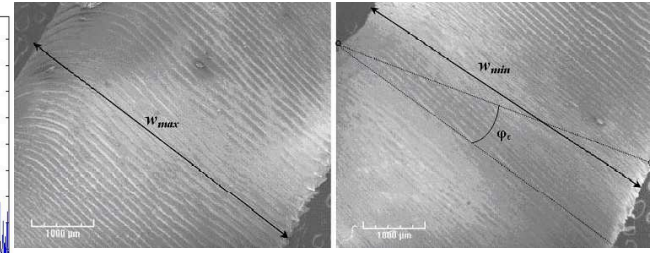}
	\caption{Chip width evaluation.}
	\label{Fig-19}
\end{figure}

\textit{The chip width} is then measured with similar techniques. Substantial width variations are observed, about $0.5~mm$. Indeed, the measured maximum width $w_{max}$ is 5.4~mm while the evaluated minimal width $w_{min}$ is 4.9~mm, which means a width variation about 10\% - Fig.~\ref{Fig-19}.

The chip width slopes angle $\varphi_{c}$ measurement (Fig.~\ref{Fig-10}) between each undulation, close to 29$^{\circ}$, is equal, with the same errors, to the phase difference measured on the signals of tool tip displacements (28$^{\circ}$) (cf. \cite{bisu-07}).

\section{Correlation between displacements of the tool tip / applied forces}
\label{CorrElationDEplacementsForces}

A synthesis between the work-paper two parts is essential. It is carried out below in order to put in evidence the various correlations which exist between stiffness / displacements, displacements of tool / stiffness center or stiffness center / central axis.

\subsection{Correlation between the plane of the displacements tool tip / the applied forces}
\label{CorrElationPlanDesDEplacementsPlanDesForces}

The tool tip point displacements are localized (cf. \cite{bisu-AAAAA-ispas-07a}) in a tilted plane. Inside the stiffness matrix determined in the paper (\cite{bisu-AA-k'nevez-07}), a correlation exists between tool tip displacements and the cutting forces applied. In particular, the ratios between the large and the small ellipses axes of tool displacements ($a_{u}$/$b_{u}$) and the efforts applied ($a_{f}$/$b_{f}$) decrease feed rate functions while the ratio of these ratios remains constant (equal to 1,64) when the feed rate increase. These elements enable to determine accurately the real configuration of cutting process. These correlations are analyzed using the direct normal to the tool tip point displacements plane and the direct normal to the load application points place (Table \ref{tabl-4}).

\begin{table*}[h!tbp]
	\centering
		\begin{tabular}{|c|c|c|c|c|c|c|c|c|}
\hline
f(mm/rev)&\multicolumn{2}{|c|}{0.05}&\multicolumn{2}{|c|}{0.0625}&\multicolumn{2}{|c|}{0.075}&\multicolumn{2}{|c|}{0.1}\\
\hline
Normal&$\vec{n_{f}}$&$\vec{n_{u}}$&$\vec{n_{f}}$&$\vec{n_{u}}$&$\vec{n_{f}}$&$\vec{n_{u}}$&$\vec{n_{f}}$&$\vec{n_{u}}$\\
\hline
Along x &0.245&-0.071&0.292&-0.071&0.419&-0.058&0.46&-0.056\\
\hline
Along y &-0.107&-0.186&-0.113&-0.186&-0.097&-0.206&-0.1&-0.216\\
\hline
Along z &-0.964&0.98&-0.95&0.98&-0.903&0.997&-0.882&0.975\\
\hline
		\end{tabular}
\caption{The normals ($\vec{n_{f}}$,$\vec{n_{u}}$) of the tool points displacements planes and the forces applied along the three cutting directions, depending on feed rate.}
	\label{tabl-4}
\end{table*}

The existence of these two planes is particularly interesting and adapted to the establishment of a cutting process real configuration. This new aspect is in the course of implementation in order to express and exploit a simplified dynamic three-dimensional model in the reference system associated with these planes \cite{bisu-AAAAA-ispas-07}.

\subsection{Self-sustained vibrations: experimental validation}
\label{sec:VibrationsAutoEntretenuesValidationExpErimentale}

\begin{figure}[htbp]
	\centering
		\includegraphics[width=0.48\textwidth]{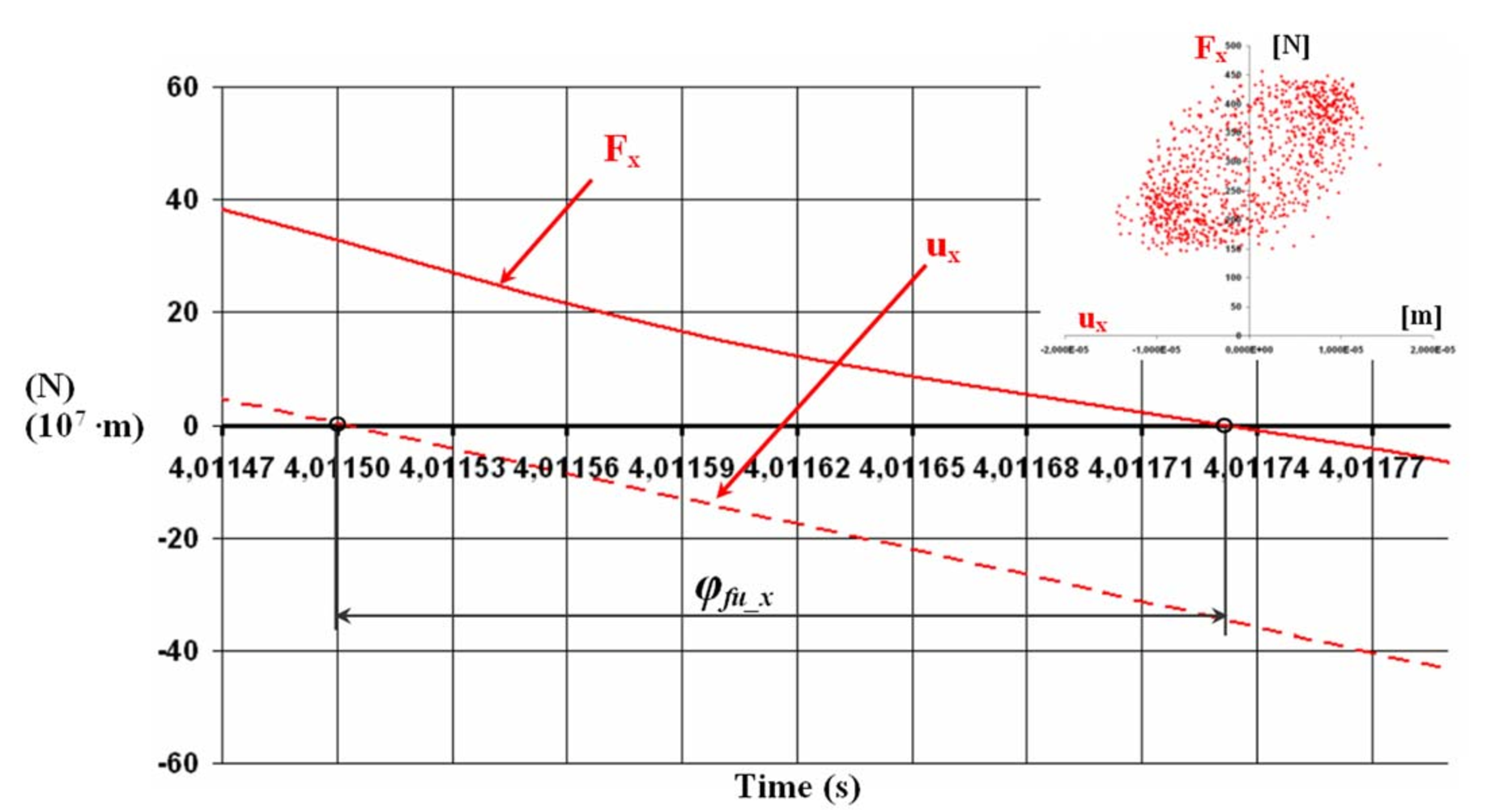}
	\caption{Phase difference evaluation between forces / displacements along x axis.}
	\label{Fig-20}
\end{figure}

\begin{figure}[htbp]
	\centering
		\includegraphics[width=0.48\textwidth]{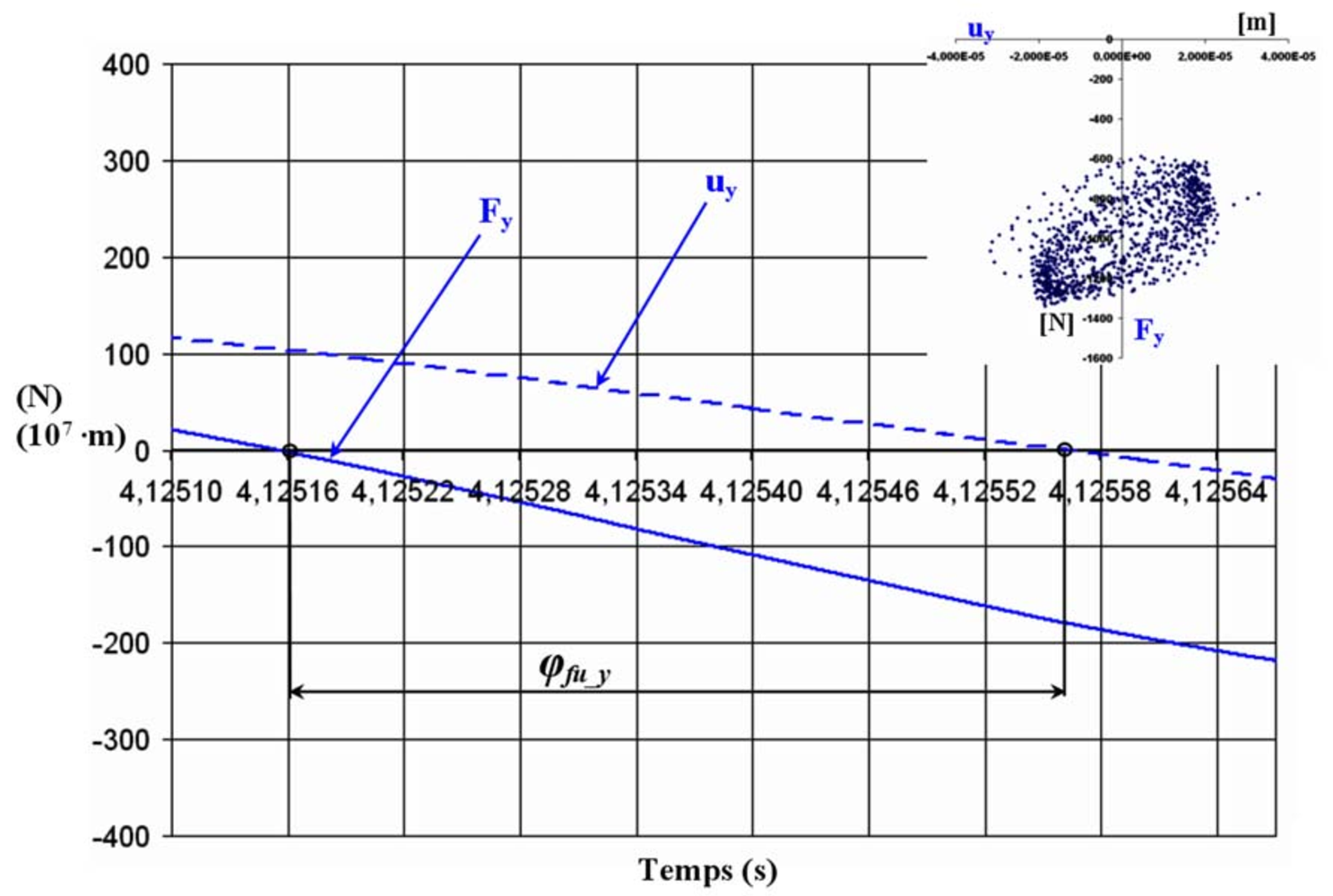}
	\caption{Phase difference evaluation between forces / displacements along y axis.}
	\label{Fig-21}
\end{figure}

\begin{figure}[htbp]
	\centering
		\includegraphics[width=0.48\textwidth]{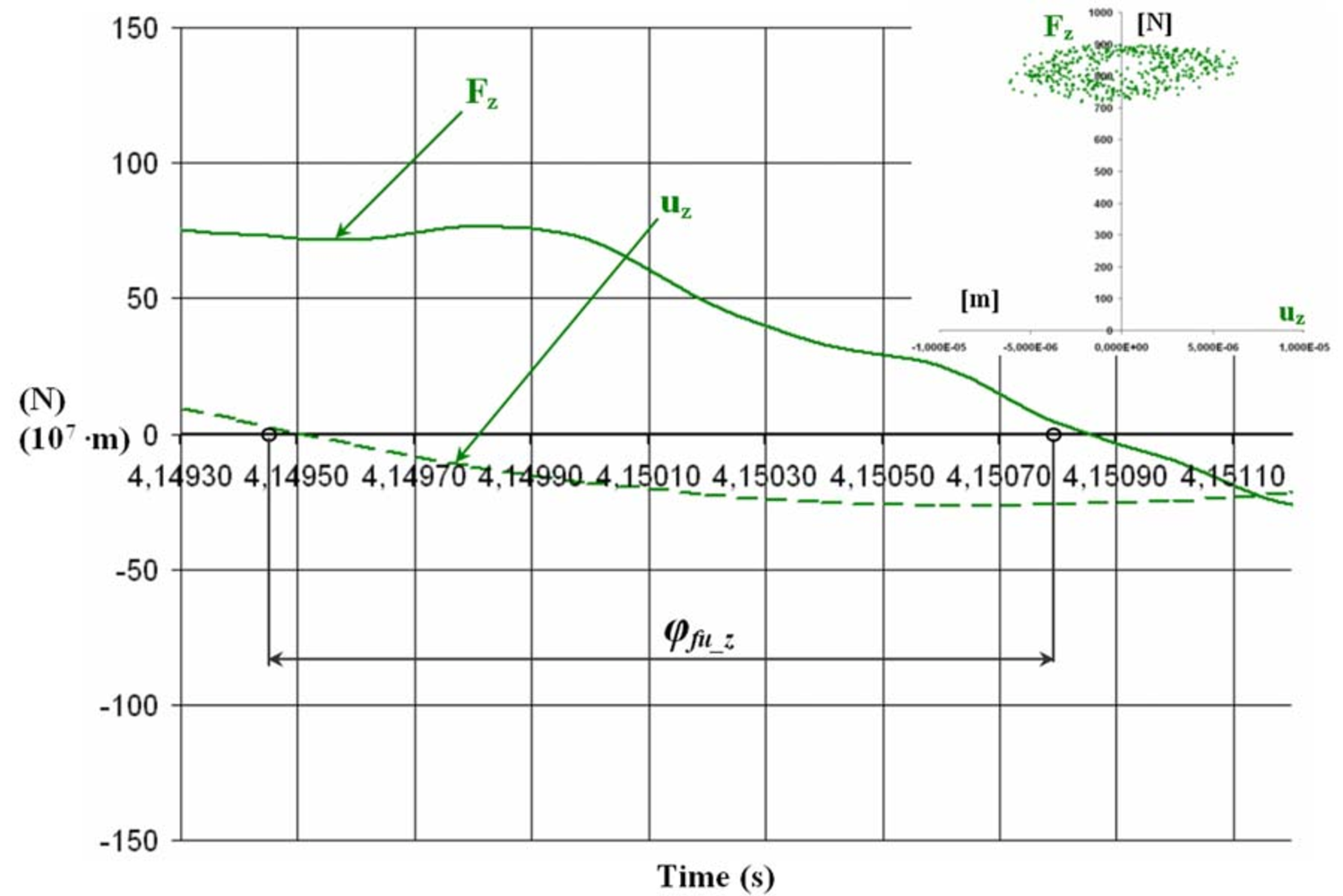}
	\caption{Phase difference evaluation between forces / displacements along z axis.}
	\label{Fig-22}
\end{figure}

From these studies, it comes out that the self-excited vibrations domain is around 190~Hz with accuracy of a few per cent. It is around this common fundamental frequency that the whole of the major characteristics (displacements, efforts, and moments) of system \textbf{WTM} have the most important amplitude variations. The analysis carried out to the measures of tool point displacements and the points for load application enables to evaluate existing constant phase difference between the forces components and corresponding displacements (Fig.~\ref{Fig-20} to \ref{Fig-22}).

These phase differences confirm the efforts delay compared to the tool tip displacement. The self-excited vibrations appearance can be also explained by the delay force / displacement, which increases the system energy level. The existence of this delay could be explained by the machining system inertia and more particularly by the cutting process inertia \cite{koenigsberger-tlusty-70}. 

\begin{table}[htbp]
\centering
		\begin{tabular}{|c|c|c|}
\hline
 $\varphi_{fu_x}$ & $\varphi_{fu_y}$ & $\varphi_{fu_z}$ \\
\hline
13$^{\circ}$ & 23$^{\circ}$ & 75$^{\circ}$ \\
\hline
\end{tabular}
\caption{Values of phase difference forces / displacements (ap = 5~mm, f = 0,1~mm/rev, N = 690~rpm.)}
	\label{tabl-5}
\end{table}

Moreover, it is noted that the phase difference with the same range between efforts / displacements (Table~\ref{tabl-5}) remains constant according to the feed rate.

Because of parts elasticity intervening in the operation of turning, it is logical that the response in displacements of unit \textbf{BT}-\textbf{BW} is carried out with a certain shift compared to the efforts variation applied to the tool, variation induced by the lacks of machined surface circularity in the preceding turn which imply variations of the contact tool/workpiece (Fig.~\ref{Fig-17}). Phase difference between the efforts and displacements thus remains a possible explanation to the self-excited vibrations appearance.

\begin{figure}[htbp]
	\centering
		\includegraphics[width=0.40\textwidth]{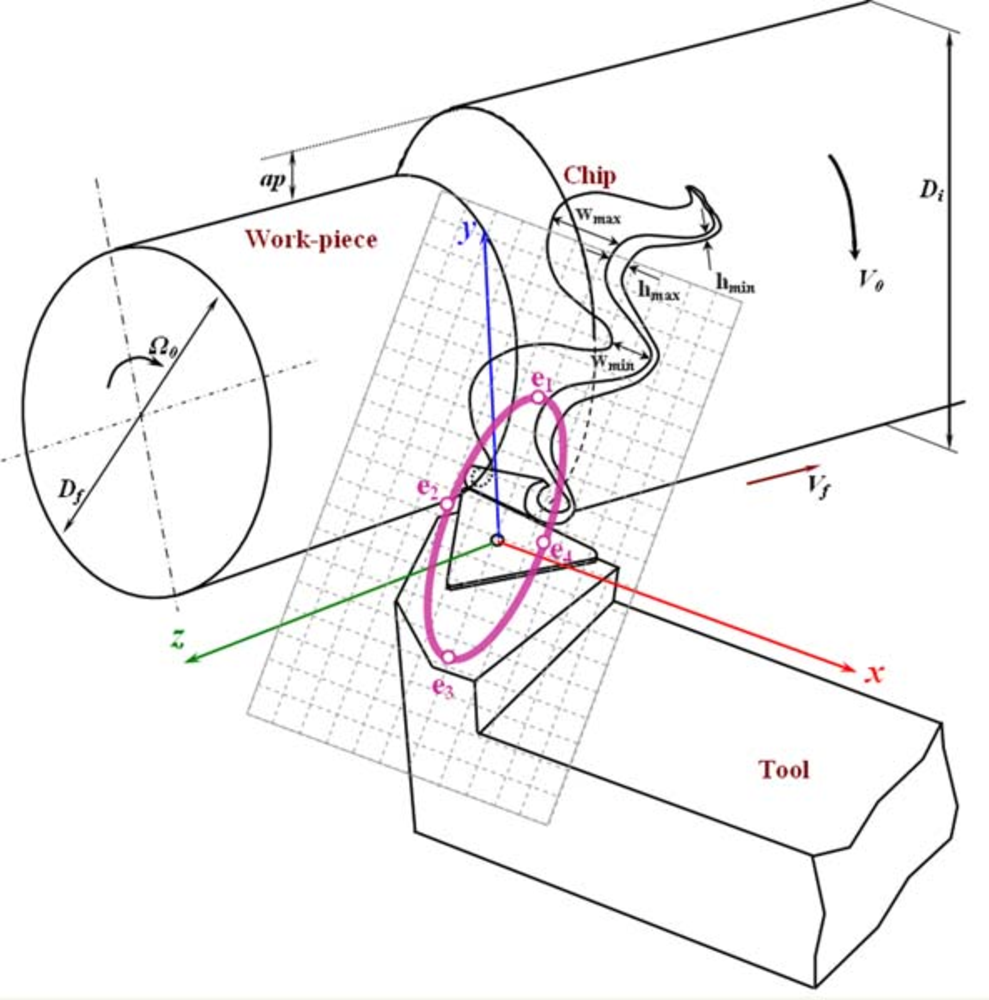}
	\caption{Movement tool / workpiece elliptical trajectory.}
	\label{Fig-23}
\end{figure}

Moreover, when the tool moves along the ellipse places $e_{1}$, $e_{2}$, $e_{3}$ (Fig.~\ref{Fig-23}), the cutting force carries out a positive work because its direction coincides with the cutting direction. On the other hand, on the side $e_{3}$, $e_{4}$, $e_{1}$, the work produced by the cutting force is negative since its direction is directly opposed to that of displacement. The comparison between these two ellipse parts, as shows that the effort on the trajectory $e_{1}$, $e_{2}$, $e_{3}$ is higher as on the trajectory $e_{3}$, $e_{4}$, $e_{1}$ because the cutting depth is more important. At the time of this process, work corresponding to an ellipse trajectory remains positive and the increase of result energy thus contributes to maintain the vibrations and to dissipate the energy in the form of heat by the assembly tool/workpiece.

\section{Conclusion}
\label{conclusion}

The experimental procedures proposed in this work-paper, as well at the static and dynamic level, enabled to determine the elements necessary to a rigorous analysis of the tool geometry influence, its displacement and evolution of the contacts tool/workpiece and tool/chip on the machined surface.

In particular, by analyzing the efforts resultant torsor applied during turning process, the experimental results enabled to establish an efforts vector decomposition highlighting the evolution of a variable cutting force around a constant value. This variable effort evolves into a plane inclined compared to the machine tool reference system.

This cutting force, whose application point describes an ellipse, is perfectly well correlated with the tip tool displacement which takes place under similar conditions. In particular, the ellipses axes ratios remains in a constant ratio when the feed rate varies with a proportionality factor equal with 1,64.

Moreover, the highlighted coupling between the system elastic characteristics \textbf{BT} and the vibrations generated by cutting enabled to demonstrate that the self-excited vibrations appearance is strongly influenced, as it was expect to, by the system stiffness, their ratio and their direction. We also established a correlation between the vibratory movement direction of the machine tool elastic structure, the thickness variations and the chip section.

These results enable to now consider a more complete study by completely exploiting the concept of torsor. Indeed, thanks to the six components dynamometer, we confirmed, for an turning operation, the moments existence to the tool tip not evaluated until now by the traditional measuring equipment. 

The originality of this work is multiple and in particular consists in a first mechanical actions torsor analysis applied to the tool tip, with an aim of making evolve a model semi analytical cutting 3D. This study allows, considering a turning operation, to establish strong correlations between the self-excited vibrations and the mechanical actions torsor central axis. It is thus possible, thanks to the parameters use defining the central axis, to study the vibrating system tool/workpiece evolution. It also leads to the description of a "plane of tool tip displacements" in correspondence with "the load application points plane".

Thus, using the plane that characterizes the \textbf{BT} behaviour enable to bring back the three dimensional cutting problem, with space displacements, with a simpler model situated in an inclined plane compared with the reference system of machine tool. Nevertheless, that remains a specific model of three dimensional cutting.
  
\section{Appendix}
\label{sec:Annexe}

\subsection{Determination of the place points plane of load application on the tool}
\label{sec:DEterminationDuPlanLieuDesPointsDApplicationDesEffortsSurLOutil}

The plane $P_{u}$ definition being load application points geometrical place on the tool starting from the experimental results is carried out using the computation Mathcad$^{\copyright}$ software. We seek to determine the plane which passes by the load application points cloud on the tool (Fig.~\ref{Fig-8} section \ref{DEcompositionPlanDesForces}):

\begin{eqnarray*}
	ax + by + cz + d = 0.
\end{eqnarray*}

The errors are noted with $e_{rr}$ and we have :

\begin{eqnarray*}
	e_{rr}(x, y,z, x_{p}, y_{p}, z_{p}, x_{n}, y_{n}, z_{n}) = \\
	\left[M(x, y, z) - P(x_{p},y_{p}, z_{p}).n_{f}(x_{n}, y_{n}, z_{n})\right] ,
\end{eqnarray*}

where:

\begin{eqnarray}
	M = \left\{x, y, z\right\}^{t}, P\left\{x_{p}, y_{p}, z_{p}\right\}^{t}, n_{f} = \left\{x_{n}, y_{n}, z_{n}\right\}^{t}.
\end{eqnarray}

Here the superscript t indicates the operation of transposition.

Expressing the $E_{rr}$ function using $e_{rr}$ and introducing the displacements components $(u_{x_{i}}, u_{y_{i}}, u_{z_{i}})$ of the tool load application points into the three space directions, it comes:

\begin{eqnarray*}
	E_{rr}(x, y,z, x_{p}, y_{p}, z_{p}, x_{n}, y_{n}, z_{n}) = \\ \sum^{N}_{i=0}e_{rr}(u_{x_{i}}, u_{y_{i}},u_{z_{i}}, x_{p}, y_{p}, z_{p}, x_{n}, y_{n}, z_{n}),
\end{eqnarray*}

and : 

\begin{eqnarray*}
	x_{p}=\frac{\sum^{N}_{i=0}u_{x_{i}}}{N}, y_{p}=\frac{\sum^{N}_{i=0}u_{y_{i}}}{N}, z_{p}=\frac{\sum^{N}_{i=0}u_{z_{i}}}{N}.
\end{eqnarray*}

Now, the vector $\vec{V}$ is calculated by minimization:

\begin{eqnarray*}
	\vec{V} = min (E_{rr}, x_{p}, y_{p},z_{p}, x_{n}, y_{n}, z_{n}) ,
\end{eqnarray*}
 
where

\begin{eqnarray*}
	\vec{V} = \left\{
	\begin{array}{c}
	V_{1} = x_{p} \\
	V_{2} = y_{p} \\
	V_{3} = z_{p} .\\
	V_{4} = x_{n} \\
	V_{5} = y_{n} \\
	V_{6} = z_{n} 
	
\end{array}\right.
\end{eqnarray*}

It results from it that the direct normal $n_{f}$ components to the plane of the load application points on the tool are given by:

\begin{eqnarray*}
	\vec{n_{f}} = \left\{
	\begin{array}{c}
	
	V_{4} \\
	V_{5} .\\
	V_{6} 
	
\end{array}\right.
\end{eqnarray*}

For the case study presented in the section \ref{DEcompositionPlanDesForces} (Table~\ref{tabl-2}), with ap = 5~mm and f= 0.1~mm it comes:

\begin{eqnarray*}
	\vec{n_{f}} = \left\{
	\begin{array}{c}
	
	+ 0.46 \\
	- 0.1 .\\
	- 0.882 
	
\end{array}\right.
\end{eqnarray*}

The required plane equation is then:

\begin{eqnarray*}
	P_{f}(s, t) = V_{p} + s.u_{1} + t.u_{2},
\end{eqnarray*}

where s and t are constants, and $\vec{V_{p}}$ is equal to :

\begin{eqnarray*}
	\vec{V_{p}} = \left\{
	\begin{array}{c}
	
	\vec{V_{1}} \\
	\vec{V_{2}} ,\\
	\vec{V_{3}} 
	
\end{array}\right.
\end{eqnarray*}

and $\vec{u_{1}}$ is given by:

\begin{eqnarray*}
	\vec{u_{1}} = \frac{\vec{u_{0} }\wedge \vec{n_{f}}}{\left\|\vec{n_{0}} \wedge \vec{u_{f}}\right\|},  \vec{u_{2}} = \vec{n }\wedge \vec{u_{1} } ,
\end{eqnarray*}

with the plane $\vec{u_{0}}$ orientation vector is:

\begin{eqnarray*}
	\vec{u_{0}} = \left\{
	\begin{array}{c}
	
	1 \\
1 .\\
	1 
	
\end{array}\right.
\end{eqnarray*} 

\subsection{Ellipse approximation}
\label{sec:ApproximationDeLEllipse}

Using the ellipse plane determination \cite{bisu-AA-knevez-08}, it is possible to determine the characteristics given in Table~\ref{tabl-3}.

\begin{acknowledgements}
The authors would like to thank the CNRS (Centre National de la Recherche Scientifique UMR 5469) for the financial support to accomplish the project. 

\end{acknowledgements}

\end{document}